\documentclass[aps,prd,reprint,preprintnumbers,groupedaddress,nofootinbib]{revtex4-2}

\usepackage[svgnames]{xcolor} 

\usepackage[printonlyused,withpage]{acronym} 
\usepackage{amsfonts}
\usepackage{anyfontsize}
\usepackage{bbm}
\usepackage{booktabs} 
\usepackage{braket}
\usepackage{cancel}
\usepackage{changepage}
\usepackage{colortbl}
\usepackage{enumitem}
\usepackage[mathscr]{euscript}
\usepackage{float}
\usepackage[T1]{fontenc} 
\usepackage{gensymb}
\usepackage[hidelinks]{hyperref} 
\usepackage{listings} 
\usepackage{makecell}
\usepackage{mathtools}
\usepackage[framemethod=tikz]{mdframed} 
\usepackage{microtype} 
\usepackage{multirow}
\usepackage{physics}
\usepackage{relsize}
\usepackage{rotating} 
\usepackage{sansmath}
\usepackage{simplewick}
\usepackage{slashed}
\usepackage{stmaryrd}
\usepackage{subfig}
\usepackage{tcolorbox}
\usepackage{theorem}
\usepackage{tikz}
\usepackage{tikz-feynman,contour}
\usepackage[normalem]{ulem}
\usepackage{xspace} 
\usepackage{yfonts}
\usepackage[vcentermath]{youngtab}

\usepackage{newtxtext,newtxmath}

\usepackage[english]{babel}
\usepackage{blindtext}

\tcbuselibrary{theorems}
\usetikzlibrary{decorations.markings}

\def\ii{{\text{i}}}

\makeatletter
\newcommand*{\transpose}{%
  {\mathpalette\@transpose{}}%
}
\newcommand*{\@transpose}[2]{%
  \raisebox{\depth}{$\m@th#1\intercal$}%
}
\makeatother

\newtcbox{\sln}{colback=Gainsboro,
colframe=Gainsboro}

\newcommand{\bt}[1]{{\sansmath{\boldsymbol{#1}}}}

\newcommand{\overbar}[1]{\mkern 2mu\overline{\mkern-4mu#1\mkern-4mu}\mkern 2mu}

\tikzset{snake it/.style={decorate, decoration={snake,amplitude=10mm}}}

\tikzset{/pgf/decoration/.cd,
    number of sines/.initial=10,
    angle step/.initial=20,
}

\newdimen\tmpdimen

\pgfdeclaredecoration{full sines}{initial}
{
    \state{initial}[
        width=+0pt,
        next state=move,
        persistent precomputation={
            \pgfmathparse{\pgfkeysvalueof{/pgf/decoration/angle step}}%
            \let\anglestep=\pgfmathresult%
            \let\currentangle=\pgfmathresult%
            \pgfmathsetlengthmacro{\pointsperanglestep}%
                {(\pgfdecoratedremainingdistance/\pgfkeysvalueof{/pgf/decoration/number of sines})/360*\anglestep}%
        }] {}
    \state{move}[width=+\pointsperanglestep, next state=draw]{
        \pgfpathmoveto{\pgfpointorigin}
    }
    \state{draw}[width=+\pointsperanglestep, switch if less than=1.25*\pointsperanglestep to final, 
        persistent postcomputation={
        \pgfmathparse{mod(\currentangle+\anglestep, 360)}%
        \let\currentangle=\pgfmathresult%
    }]{%
        \pgfmathsin{+\currentangle}%
        \tmpdimen=\pgfdecorationsegmentamplitude%
        \tmpdimen=\pgfmathresult\tmpdimen%
        \divide\tmpdimen by2\relax%
        \pgfpathlineto{\pgfqpoint{0pt}{\tmpdimen}}%
    }
    \state{final}{
        \ifdim\pgfdecoratedremainingdistance>0pt\relax
            \pgfpathlineto{\pgfpointdecoratedpathlast}
        \fi
   }
}

\pgfdeclaredecoration{completely sines}{initial}
{
    \state{initial}[
        width=+0pt,
        next state=upsine,
        persistent precomputation={\pgfmathsetmacro\matchinglength{
            \pgfdecoratedinputsegmentlength / int(\pgfdecoratedinputsegmentlength/\pgfdecorationsegmentlength)}
            \setlength{\pgfdecorationsegmentlength}{\matchinglength pt}
        }] {}
    \state{upsine}[width=\pgfdecorationsegmentlength,next state=downsine]{
        \pgfpathsine{\pgfpoint{0.25\pgfdecorationsegmentlength}{0.5\pgfdecorationsegmentamplitude}}
        \pgfpathcosine{\pgfpoint{0.25\pgfdecorationsegmentlength}{-0.5\pgfdecorationsegmentamplitude}}
    }
    \state{downsine}[width=\pgfdecorationsegmentlength,next state=upsine]{
        \pgfpathsine{\pgfpoint{0.25\pgfdecorationsegmentlength}{-0.5\pgfdecorationsegmentamplitude}}
        \pgfpathcosine{\pgfpoint{0.25\pgfdecorationsegmentlength}{0.5\pgfdecorationsegmentamplitude}}
}
    \state{final}{}
}

\tikzfeynmanset{ with arrow/.style = {
   decoration={
     markings,
     mark=at position 0.5
          with {\arrow[xshift=1.9mm]{Latex[width=2.75mm,length=3mm]}}
     },
   postaction=decorate}
}

\tikzfeynmanset{ with glarrow/.style = {
   decoration={
     markings,
     mark=at position 0.5
          with {\arrow[white,xshift=3mm]{Latex[width=4.125mm,length=4.5mm]}}
     },
   postaction=decorate}
}

\tikzfeynmanset{ with outarrow/.style = {
   decoration={
     markings,
     mark=at position 0.5
          with {\arrow[xshift=3.35mm]{Latex[width=4.58333mm,length=5mm]}}
     },
   postaction=decorate}
}

\tikzfeynmanset{ bigphoton/.style={
    /tikz/draw=none,
    /tikz/decoration={name=none},
    /tikz/postaction={
      /tikz/draw,
      /tikz/decoration={
        completely sines,
        segment length=3mm,
        amplitude=2.5mm,
      },
      /tikz/decorate=true,
    },
  },
}

\tikzfeynmanset{ roundbigphoton/.style={
    /tikz/draw=none,
    /tikz/decoration={name=none},
    /tikz/postaction={
      /tikz/draw,
      /tikz/decoration={
        full sines,
        segment length=3mm,
        amplitude=1.25mm,
      },
      /tikz/decorate=true,
    },
  },
}

\tikzset{ mega thick/.style= {line width = 3.4pt}
}

\makeatletter
\renewcommand{\fnum@figure}{\textsc{\figurename~\thefigure}} 
\makeatother

\lstnewenvironment{pseudoc}{\lstset{frame=lines,language=C,mathescape=true}}{}
\lstnewenvironment{logs}{\lstset{frame=lines,basicstyle=\footnotesize\ttfamily,numbers=none}}{}
\lstnewenvironment{cc}{\lstset{frame=lines,language=C}}{}
\lstnewenvironment{c64}{\lstset{backgroundcolor=\color{c64},basewidth=0.65em,basicstyle=\commodoreface\color{c64light},numbers=none,framerule=10pt,rulecolor=\color{c64light},frame=tb,framexbottommargin=30pt}}{}
\lstnewenvironment{html}{\lstset{frame=lines,language=html,numbers=none}}{}
\lstnewenvironment{pseudo}{\lstset{frame=lines,mathescape=true,morekeywords={learn_string_domain, save_model}}}{}
\lstnewenvironment{pseudoctiny}{\lstset{language=C,mathescape=true,basicstyle=\tiny\sffamily}}{}
\lstnewenvironment{cctiny}{\lstset{language=C,basicstyle=\tiny\sffamily}}{}
\lstnewenvironment{pseudotiny}{\lstset{mathescape=true,basicstyle=\tiny\sffamily}}{}

\begin{document}

\title{The phenomenological cornucopia of SU(3) exotica}

\preprint{UCI-HEP-TR-2021-26}

\author{Linda M. Carpenter}
\email{lmc@physics.osu.edu}
\author{Taylor Murphy}
\email{murphy.1573@osu.edu}
\affiliation{Department of Physics, The Ohio State University\\ 191 W. Woodruff Ave., Columbus, OH 43212, U.S.A.}
\author{Tim M. P. Tait}
\email{ttait@uci.edu}
\affiliation{Department of Physics and Astronomy, University of California, Irvine\\ Irvine, CA 92697, U.S.A.}

\date{\today}

\begin{abstract}
We introduce an effort to catalog the gauge-invariant interactions of Standard Model (SM) particles and new fields in a variety of representations of the SM color gauge group $\mathrm{SU}(3)_{\text{c}}$. In this first installment, we direct this effort toward fields in the six-dimensional (sextet, $\boldsymbol{6}$) representation. We consider effective operators of mass dimension up to seven (comprehensively up to dimension six), featuring both scalar and fermionic color sextets. We use an iterative tensor-product method to identify the color invariants underpinning such operators, emphasizing structures that have received little attention to date. In order to demonstrate the utility of our approach, we study a simple but novel model of color-sextet fields at the Large Hadron Collider (LHC). We compute cross sections for an array of new production channels enabled by our operators, including single-sextet production and sextet production in association with photons or leptons. We also discuss dijet-resonance constraints on a sextet fermion. This example shows that there remains a wide array of fairly minimal but well motivated and unexplored models with extended strong sectors as we await the high-luminosity LHC.
\end{abstract}

\maketitle

\section{Introduction}
\label{s1}

The search for physics beyond the Standard Model (bSM), spearheaded by the CERN Large Hadron Collider (LHC) \cite{Morrissey:2009tf}, continues unabated, spurred on by a constellation of theoretical deficiencies and experimental anomalies that continue to bedevil the Standard Model. These issues range from the well known naturalness problems \cite{PhysRevLett.37.8,tHooft:1979rat,PhysRevD.88.098301,deGouvea:2014xba} to the collection of persistent muon \cite{PhysRevLett.126.141801} and flavor anomalies \cite{Aaij_2014,lhcbcollaboration2021test} and the ongoing search for particle dark matter \cite{Lee:1977ua,Bertone_2018}. While there exist a wide variety of specific, sometimes ultraviolet-complete, frameworks that seek to rectify some or all of the Standard Model's shortcomings, some --- in the absence of experimental evidence for any specific bSM model --- have turned to an effective field theory (EFT) approach to study new physics in a simplified and more model-independent manner.

The EFT approach, which in the simplest terms allows one to study the experimentally accessible degrees of freedom in a theory while remaining agnostic about physics at higher energy scales, has been used to great effect in many contexts. Some pertinent examples include the Standard Model Effective Field Theory (SMEFT) \cite{Grzadkowski_2010}, which now includes (in at least one basis) every independent operator comprising SM fields of up to mass dimension eight \cite{Brivio_2017,Murphy_2020,another_dim_8} and has lately been used to probe experimental anomalies; and the variety of effective operators employed in supersymmetric frameworks in which supersymmetry breaking is communicated by heavy messengers from a hidden sector to the visible world \cite{Fox:2002bu,Carpenter:2015mna}.

Lying somewhere between the SMEFT and supersymmetric models on the scale of bSM theories ranked by exotic particle content are 
simplified models in which the SM is augmented only by new matter whose only non-trivial gauge transformations are under the SM gauge group $\mathcal{G}_{\text{SM}} = \mathrm{SU}(3)_{\text{c}} \times \mathrm{SU}(2)_{\text{L}} \times \mathrm{U}(1)_Y$. A particularly simple but fruitful subset of these models feature a new color-charged $\mathrm{SU}(2)_{\text{L}}$ singlet perhaps with non-vanishing weak hypercharge. While new color-charged fields --- particularly $\mathrm{SU}(3)_{\text{c}}$ triplets and octets (adjoints) --- occur in a panoply of bSM theories and have long been studied in those specific contexts, there has been far less attempt at a coherent and comprehensive accounting of bSM strong interactions in an EFT framework. We intend to 
address this gap in the literature, beginning with the present work.

We construct a catalog of effective operators governing fields in the six-dimensional (sextet) representation of $\mathrm{SU}(3)_{\text{c}}$, which remain
hypothetical but can be copiously produced in proton-proton ($pp$) collisions and are therefore highly relevant to the ongoing LHC program. Using an iterative tensor-product method to construct new color invariants, we enumerate all possible operators of mass dimension up to six connecting sextets to Standard Model fields, and additionally identify a few potentially interesting dimension-seven operators. Several of these operators have to date received little or no attention. In order to motivate this operator catalog \emph{ex post facto}, we construct two simplified models of color-sextet fields and explore the phenomenology of these particles at the LHC. These models are notably different from previous models of color sextets in that the novel particles couple not to quark pairs, like the aptly named ``sextet diquarks'' \cite{Shu:2009xf,Han_2010,Han_2010_2}, but instead to a quark and a gluon. We find fairly light constraints on sextet fermions from a CMS search for dijet resonances \cite{CMS:2018s1} and propose future searches for the other interesting multijet + $\gamma$/$Z$ or lepton signatures generated by these simple models.

This paper is organized as follows. In \hyperref[s2]{Section II}, we show how to build a catalog of exotic-color operators starting from first principles, and we produce such a catalog for color-sextet fields. In \hyperref[s3]{Section III}, we use a small subset of these operators to explore a simplified model of color-sextet scalars and fermions. We compute cross sections for a variety of production modes and discuss LHC phenomenology while surveying constraints on these novel particles. \hyperref[s4]{Section IV} concludes. The \hyperref[aA]{appendix} contains a thorough discussion of the representation theory of the $\mathrm{SU}(3)$ sextet, along with some notes about the implementation of our specific color-sextet models in public computing tools.
\section{Effective interactions of exotic color-charged states}
\label{s2}

Our aim is to catalog the interactions of new $\mathrm{SU}(3)_{\text{c}}$-charged matter with the Standard Model. Ultimately, one could imagine
considering bSM fields up to the twenty-seven-dimensional representation of $\mathrm{SU}(3)_{\text{c}}$, which is
the highest representation that can be produced resonantly in a $pp$ collision \emph{via} $gg$ fusion. In order to generate results that are novel and useful but also relatively simple, we choose to focus on particles in the six-dimensional (sextet, $\boldsymbol{6}$) representation of $\mathrm{SU}(3)_{\text{c}}$. The method we describe can be easily generalized to other color representations.

\subsection{Color singlets by iteration}
\label{s2.1}

The goal is to find all gauge- and Lorentz-invariant operators governing exotic color-charged matter and the Standard Model. As an initial step, we need to identify all relevant\footnote{At present, our goal is a comprehensive catalog at dimensions five and six, but we identify interesting dimension-seven operators throughout. In other words, we ignore $\mathrm{SU}(3)_{\text{c}}$ invariants that can only produce dimension-eight or higher (smaller) operators.} gauge-invariant contractions of SM color-charged fields with color sextets. We therefore begin by enumerating the gauge-invariant contractions of three color-charged fields that can be realized at the LHC. To do this, we recall the tensor decompositions of direct product (reducible) representations $N \otimes M$, $\{N,M\} \leq \boldsymbol{8}$, of $\mathrm{SU}(3)$  \cite{Slansky:1981yr,Georgi:1982jb}:
\begin{align}\label{decomp}
\nonumber     \boldsymbol{3} \otimes \boldsymbol{3} &= \boldsymbol{\bar{3}}_{\text{a}} \oplus \boldsymbol{6}_{\text{s}},\\
\nonumber    \boldsymbol{3} \otimes \boldsymbol{\bar{3}} &= \boldsymbol{1} \oplus \boldsymbol{8},\\
\nonumber    \boldsymbol{6} \otimes \boldsymbol{3} &= \boldsymbol{8} \oplus \boldsymbol{10},\\
\nonumber    \boldsymbol{6} \otimes \boldsymbol{\bar{3}} &= \boldsymbol{3} \oplus \boldsymbol{15},\\
\nonumber    \boldsymbol{6} \otimes \boldsymbol{6} &= \boldsymbol{\bar{6}}_{\text{s}} \oplus \boldsymbol{15}_{\text{a}} \oplus \boldsymbol{15'}_{\!\!\text{s}},\\
\nonumber    \boldsymbol{6} \otimes \boldsymbol{\bar{6}} &= \boldsymbol{1} \oplus \boldsymbol{8} \oplus \boldsymbol{27},\\
\nonumber    \boldsymbol{8} \otimes \boldsymbol{3} &= \boldsymbol{3} \oplus \boldsymbol{\bar{6}} \oplus \boldsymbol{15},\\
\nonumber    \boldsymbol{8} \otimes \boldsymbol{\bar{6}} &= \boldsymbol{3} \oplus \boldsymbol{\bar{6}} \oplus \boldsymbol{15} \oplus \boldsymbol{24},\\
\text{and}\ \ \    \boldsymbol{8} \otimes \boldsymbol{8} &= \boldsymbol{1}_{\text{s}} \oplus \boldsymbol{8}_{\text{s}} \oplus \boldsymbol{8}_{\text{a}} \oplus \boldsymbol{10}_{\text{a}} \oplus \boldsymbol{\overbar{10}}_{\, \text{a}} \oplus \boldsymbol{27}_{\text{s}},
\end{align}
where the subscripts \textbf{s} (\textbf{a}) indicate a symmetric (antisymmetric) contraction. There are four reducible representations of $\mathrm{SU}(3)$, given by the direct product of three irreducible representations (smaller than $\boldsymbol{8}$ and including at least one sextet) that contain a color singlet. As a shorthand, we refer to these reducible representations as \emph{invariants}, and we write them as
\begin{align}\label{three_field_invariants}
\nonumber &\boldsymbol{3} \otimes \boldsymbol{3} \otimes \boldsymbol{\bar{6}},\\
\nonumber &\boldsymbol{3} \otimes \boldsymbol{6} \otimes \boldsymbol{8},\\
\nonumber &\boldsymbol{6} \otimes \boldsymbol{6} \otimes \boldsymbol{6},\\
\text{and}\ \ \ &\boldsymbol{6} \otimes \boldsymbol{\bar{6}} \otimes \boldsymbol{8}.
\end{align}
This shorthand indicates that there exists (at least) one way to contract fields in the corresponding representations of $\mathrm{SU}(3)_{\text{c}}$ that results in a color singlet.  These ``three-field'' invariants produce a number of interesting operators by themselves, but only scratch the surface of what is possible in $\mathrm{SU}(3)$. In order to go deeper, we make a straightforward observation about reducible representations of $\mathrm{SU}(3)$ based on the simple fact that the direct product of an irreducible representation $\textbf{p}$ with its conjugate $\boldsymbol{\bar{\textbf{p}}}$ contains a gauge singlet. In particular:\\

\emph{Observation}. If there exist invariant combinations of $n+1$ and $m+1$ fields transforming in the direct product representations $\textbf{r}_1 \otimes \cdots \otimes \textbf{r}_n \otimes \textbf{p}$ and $\textbf{q}_1 \otimes \cdots \otimes \textbf{q}{}_{m} \otimes \textbf{p}$ of $\mathrm{SU}(3)$, then there exists an invariant combination of $n+m$ fields in the reducible representation $\textbf{r}_1 \otimes \cdots \otimes \textbf{r}_n \otimes \boldsymbol{\bar{\textbf{q}}}_1 \otimes \cdots \otimes \boldsymbol{\bar{\textbf{q}}}{}_{m}$.\\

\noindent This observation allows us to systematically identify all gauge-invariant color structures in and beyond the Standard Model. Applying this technique to the list \eqref{decomp}, with $n=m=2$, yields ten independent invariants of four fields that can be used to construct effective operators of dimension seven or lower (larger) including at least one sextet and at least one SM field. We write these ``four-field'' invariants in order of increasing representation dimension as
\begin{align}\label{four_field_invariants}
\nonumber &\boldsymbol{3} \otimes \boldsymbol{3} \otimes \boldsymbol{6} \otimes \boldsymbol{6},\\
\nonumber &\boldsymbol{3} \otimes \boldsymbol{3} \otimes \boldsymbol{\bar{6}} \otimes \boldsymbol{8},\\
\nonumber &\boldsymbol{3} \otimes \boldsymbol{\bar{3}} \otimes \boldsymbol{\bar{3}} \otimes \boldsymbol{\bar{6}},\\
\nonumber &\boldsymbol{3} \otimes \boldsymbol{\bar{3}} \otimes \boldsymbol{6} \otimes \boldsymbol{\bar{6}},\\
\nonumber &\boldsymbol{3} \otimes \boldsymbol{6} \otimes \boldsymbol{6} \otimes \boldsymbol{\bar{6}},\\
\nonumber &\boldsymbol{3} \otimes \boldsymbol{6} \otimes \boldsymbol{8} \otimes \boldsymbol{8},\\
\nonumber &\boldsymbol{3} \otimes \boldsymbol{\bar{6}} \otimes \boldsymbol{\bar{6}} \otimes \boldsymbol{8},\\
\nonumber &\boldsymbol{6} \otimes \boldsymbol{6} \otimes \boldsymbol{\bar{6}} \otimes \boldsymbol{\bar{6}},\\
\nonumber &\boldsymbol{6} \otimes \boldsymbol{6} \otimes \boldsymbol{6} \otimes \boldsymbol{8},\\
\text{and}\ \ \ &\boldsymbol{6} \otimes \boldsymbol{\bar{6}} \otimes \boldsymbol{8} \otimes \boldsymbol{8}.
\end{align}
This larger set of invariants underpins a significantly larger set of gauge-invariant operators than what is generated by the three-field invariants. The final iteration required to fit the scope of this work takes $n=2$, $m=3$, and produces the following list of ten ``five-field'' invariants:
\begin{align}\label{five_field_invariants}
\nonumber &\boldsymbol{3} \otimes \boldsymbol{3} \otimes \boldsymbol{3} \otimes \boldsymbol{3} \otimes \boldsymbol{6},\\
\nonumber &\boldsymbol{3} \otimes \boldsymbol{3} \otimes \boldsymbol{3} \otimes \boldsymbol{\bar{3}} \otimes \boldsymbol{\bar{6}},\\
\nonumber &\boldsymbol{3} \otimes \boldsymbol{3} \otimes \boldsymbol{6} \otimes \boldsymbol{\bar{6}} \otimes \boldsymbol{\bar{6}},\\
\nonumber &\boldsymbol{3} \otimes \boldsymbol{\bar{3}} \otimes \boldsymbol{6} \otimes \boldsymbol{6} \otimes \boldsymbol{6},\\
\nonumber &\boldsymbol{3} \otimes \boldsymbol{\bar{3}} \otimes \boldsymbol{6} \otimes \boldsymbol{\bar{6}} \otimes \boldsymbol{8},\\
\nonumber &\boldsymbol{3} \otimes \boldsymbol{6} \otimes \boldsymbol{6} \otimes \boldsymbol{6} \otimes \boldsymbol{6},\\\
\nonumber &\boldsymbol{3} \otimes \boldsymbol{6} \otimes \boldsymbol{\bar{6}} \otimes \boldsymbol{\bar{6}} \otimes \boldsymbol{\bar{6}},\\
\nonumber &\boldsymbol{6} \otimes \boldsymbol{6} \otimes \boldsymbol{6} \otimes \boldsymbol{\bar{6}} \otimes \boldsymbol{\bar{6}},\\
\nonumber &\boldsymbol{6} \otimes \boldsymbol{6} \otimes \boldsymbol{\bar{6}} \otimes \boldsymbol{\bar{6}} \otimes \boldsymbol{8},\\
\text{and}\ \ \ &\boldsymbol{6} \otimes \boldsymbol{6} \otimes \boldsymbol{6} \otimes \boldsymbol{8} \otimes \boldsymbol{8}.
\end{align}
Many of the invariants in \eqref{four_field_invariants} and \eqref{five_field_invariants} produce operators with a minimum mass dimension of seven, but several work at dimension six. At any rate, the three preceding lists provide all the group-theoretic ingredients of dimension-seven or lower operators governing color-charged fields. As we implied in the \hyperref[s1]{Introduction}, we build operators from these invariants assuming that only the sextet is novel; \emph{i.e.}, we take the triplets ($\boldsymbol{3}$) to be SM quarks and the octets ($\boldsymbol{8}$) to be SM gluons. It should therefore be noted that specific assignments of sextet weak hypercharge may be necessary in order to preserve invariance under the full SM gauge group $\mathcal{G}_{\text{SM}}$.

Most of the four- and five-field invariants enumerated above permit more than one contraction, which happens when there exists more than one intermediate representation $\textbf{p}$ implying the $\mathrm{SU}(3)$ invariant $\textbf{r}_1 \otimes \cdots \otimes \textbf{r}_n \otimes \boldsymbol{\bar{\textbf{q}}}_1 \otimes \cdots \otimes \boldsymbol{\bar{\textbf{q}}}{}_m$ (as defined in the observation on the previous page). Distinct gauge-invariant contractions of a fixed set of fields are realized with distinct sets of \emph{generalized Clebsch-Gordan} coefficients. As an example, the invariant composed of three color octets can be built in a totally symmetric or a totally antisymmetric manner, and an observable associated with these two possible vertices (summed over colors) carries a factor proportional to
\begin{align}
    f_{abc}f^{abc} = 24\ \ \ \emph{vs.}\ \ \ d_{abc}\,d^{abc} = \frac{40}{3}\ \ \ \text{for $\mathrm{SU}(3)_{\text{c}}$}.
\end{align}
In this example, the Clebsch-Gordan coefficients correspond to the structure constants $f^{abc}$ and the $\mathrm{SU}(3)$ totally symmetric symbol, $d^{abc} = 2 \tr\, \{\bt{t}_{\boldsymbol{3}}^a,\bt{t}_{\boldsymbol{3}}^b\}\bt{t}^c_{\boldsymbol{3}}$ with $\bt{t}_{\textbf{r}}^a$ the generators of the irreducible representation \textbf{r} of $\mathrm{SU}(3)$. Other Clebsch-Gordan coefficients have been computed in prior studies of exotic color-sextet fields; these are the coefficients linking three color triplets ($\bt{L}^{ijk}$, totally antisymmetric) and two triplets to an antisextet ($\bt{K}_s^{\ \,ij}$, symmetric under $i\leftrightarrow j$ interchange) \cite{Han_2010}. In Tables \hyperref[t1]{I}--\hyperref[t3]{III}, we specify all possible contractions of color indices generating the invariants listed in \eqref{three_field_invariants}, \eqref{four_field_invariants}, and \eqref{five_field_invariants}. In these tables, and throughout this document, index heights in Clebsch-Gordan coefficients (associated with irreducible representations that are not self-conjugate) are fixed in order to contract correctly with fields in the representations specified in the first column, so for instance the totally antisymmetric coefficients for the invariant $\boldsymbol{3} \otimes \boldsymbol{3} \otimes \boldsymbol{3}$ are written so that
\begin{align}\label{inv_example}
\mathcal{L} \supset \bt{L}^{ijk} \varphi_i \varphi_j \varphi_k
\end{align}
would be a color singlet composed of three scalars $\varphi_i$ in the fundamental representation of $\mathrm{SU}(3)_{\text{c}}$. 
\hyperref[aA]{Appendix A} contains a more thorough discussion of Clebsch-Gordan coefficients, including further explanation of notation, a method for computing them, and some important results used later in this work.

\renewcommand*{\arraystretch}{2}
\begin{table*}[t!]\label{t1}
\begin{center}
 \begin{tabular}{|c||c|c|c|}
\hline
\rule{0pt}{3.4ex}Invariant & \multicolumn{2}{c|}{\ Clebsch-Gordan coefficients\ \ } & Notes \\[0.83ex]
\hline
\hline
\rule{0pt}{3.5ex}$\boldsymbol{3} \otimes \boldsymbol{3} \otimes \boldsymbol{3}$ & $\bt{L}^{ijk}$ & &  Totally antisymmetric \\[0.83ex]
\hline
\rule{0pt}{3.5ex}$\boldsymbol{3} \otimes \boldsymbol{3} \otimes \boldsymbol{\bar{6}}$ & $\bt{K}_s^{\ \,ij}$ & &  $i \leftrightarrow j$ symmetric \\[0.83ex]
\hline
 \rule{0pt}{3.5ex}$\boldsymbol{3} \otimes \boldsymbol{\bar{3}} \otimes \boldsymbol{8}$ & \ $[\bt{t}_{\boldsymbol{3}}^a]_i^{\ \,j}$\ \ & &  Generators of $\boldsymbol{3}$\\[0.83ex]
 \hline
 \rule{0pt}{3.5ex}$\boldsymbol{3} \otimes \boldsymbol{6} \otimes \boldsymbol{8}$ & $\bt{J}^{\,s\,ia}$ & &  See \hyperref[aA]{Appendix A}\\[0.83ex]
\hline
\rule{0pt}{3.5ex}$\boldsymbol{3} \otimes \boldsymbol{6} \otimes \boldsymbol{\overbar{10}}$ & $\bt{E}_x^{\ \,is}$ & &  \\[0.83ex]
\hline
\rule{0pt}{3.5ex}$\boldsymbol{3} \otimes \boldsymbol{\bar{6}} \otimes \boldsymbol{15}$ & $\bt{Q}^{\,q\, i}_{\ \ \  s}$ & &  \\[0.83ex]
\hline
\rule{0pt}{3.5ex}$\boldsymbol{3} \otimes \boldsymbol{8} \otimes \boldsymbol{\overbar{15}}$ & $\bt{V}_q^{\ \, ia}$ & &  \\[0.83ex]
\hline
 \rule{0pt}{3.5ex}$\boldsymbol{6} \otimes \boldsymbol{6} \otimes \boldsymbol{6}$ & $\bt{S}^{\,stu}$ & &  $t \leftrightarrow u$ symmetric \\[0.83ex]
 \hline
 \rule{0pt}{3.5ex}$\boldsymbol{6} \otimes \boldsymbol{6} \otimes \boldsymbol{\overbar{15}}$ & $\bt{W}_q^{\ \,st}$ & &  $s \leftrightarrow t$ antisymmetric \\[0.83ex]
\hline
 \rule{0pt}{3.5ex}$\boldsymbol{6} \otimes \boldsymbol{\bar{6}} \otimes \boldsymbol{8}$ & $[\bt{t}_{\boldsymbol{6}}^a]_s^{\ \,t}$ & &  Generators of $\boldsymbol{6}$ \\[0.83ex]
 \hline
  \rule{0pt}{3.5ex}$\boldsymbol{6} \otimes \boldsymbol{\bar{6}} \otimes \boldsymbol{27}$ & $\bt{F}^{\,n\ t}_{\ \,s}$ & &  \\[0.83ex]
 \hline
 \rule{0pt}{3.5ex}$\boldsymbol{6} \otimes \boldsymbol{8} \otimes \boldsymbol{15}$ & $\bt{X}^{q\, s a}$ & &  \\[0.83ex]
 \hline
  \rule{0pt}{3.5ex}$\boldsymbol{8} \otimes \boldsymbol{8} \otimes \boldsymbol{8}$ & \ \ \ \ \ \  $f^{abc}$\ \ \ \ \ \ \  & \ $d^{abc}$\ \ &  \ ($f^{abc}$) $d^{abc}$ totally (anti)symmetric\ \ \\[0.83ex]
  \hline
    \rule{0pt}{3.5ex}$\boldsymbol{8} \otimes \boldsymbol{8} \otimes \boldsymbol{10}$ & $\bt{G}^{\,x\,ab}$ &  &  \ $a\leftrightarrow b$ antisymmetric\ \ \\[0.83ex]
    \hline
      \rule{0pt}{3.5ex}\ $\boldsymbol{8} \otimes \boldsymbol{8} \otimes \boldsymbol{27}$\ \ & $\bt{H}^{\,n\,ab}$ & &  \ $a\leftrightarrow b$ symmetric\ \ \\[0.83ex]
 \hline
  \end{tabular}\\[0.83ex]
 \end{center}
 \caption{Color index contractions yielding the three-field color invariants needed for a catalog of dimension-six or lower operators governing color sextet interactions with SM quarks and gluons. This table establishes our notation for invariant (Clebsch-Gordan) coefficients, some of which are used for phenomenology later in this work. Tables \hyperref[t2]{II} and \hyperref[t3]{III} extend this catalog to four- and five-field invariants.}
 \end{table*}
\renewcommand*{\arraystretch}{1}

\renewcommand*{\arraystretch}{2}
\begin{table*}\label{t2}
\begin{center}
 \begin{tabular}{|c||c|c|c|c|c|c|c|c|c|c|}
\hline
\rule{0pt}{3.4ex}Invariant & \multicolumn{7}{c|}{\ Clebsch-Gordan coefficients\ \ } & Notes \\[0.83ex]
\hline
\hline
\rule{0pt}{3.5ex}\ $\boldsymbol{3} \otimes \boldsymbol{3} \otimes \boldsymbol{6} \otimes \boldsymbol{6}$\ \ &   \multicolumn{2}{c|}{} & $\bt{K}_{u}^{\ \,ij}\, \bt{S}^{\,u st}$ & \multicolumn{4}{c|}{} & $\ni [\Pi_{\boldsymbol{3366}}]^{ij\,st}$ \\[0.83ex]
 \hline
 \rule{0pt}{3.5ex}$\boldsymbol{3} \otimes \boldsymbol{3} \otimes \boldsymbol{\bar{6}} \otimes \boldsymbol{8}$ &  $\bt{L}^{ijk} \bar{\bt{J}}_{s\, ka}$ & $\bt{K}_s^{\ \,ik} [\bt{t}_{\boldsymbol{3}}^a]_k^{\ \,j}$ & $\bt{K}_r^{\ \,ij}[\bt{t}_{\boldsymbol{6}}^a]_s^{\ r}$ & \multicolumn{2}{c|}{} & $\bt{Q}^{\,q\, i}_{\ \ \ s}\bt{V}_q^{\ \,ja}$ & & $\ni [\Pi_{\boldsymbol{33\bar{6}8}}]^{ij\ \, a}_{\ \  s}$ \\[0.83ex]
 \hline
  \rule{0pt}{3.5ex}$\boldsymbol{3} \otimes \boldsymbol{\bar{3}} \otimes \boldsymbol{\bar{3}} \otimes \boldsymbol{\bar{6}}$ &  $\bar{\bt{L}}_{\,jkl}\,\bt{K}_s^{\ \,li}$ & \multicolumn{2}{c|}{} & $[\bt{t}_{\boldsymbol{3}}^a]_j^{\ i}\, \bar{\bt{J}}_{ s\, ak}$ & \multicolumn{3}{c|}{} & $\ni [\Pi_{\boldsymbol{3\bar{3}\bar{3}\bar{6}}}]^i_{\ jks}$ \\[0.83ex]
 \hline
   \rule{0pt}{3.5ex}$\boldsymbol{3} \otimes \boldsymbol{\bar{3}} \otimes \boldsymbol{6} \otimes \boldsymbol{\bar{6}}$ &  $\delta_j^{\ \,i}\,\delta_{t}^{\ \,s}$ &\multicolumn{2}{c|}{} & $[\bt{t}_{\boldsymbol{3}}^a]_j^{\ \, i}[\bt{t}_{\boldsymbol{6}}^a]_t^{\ \, s}$ & \multicolumn{3}{c|}{} & $\ni [\Pi_{\boldsymbol{3\bar{3}6\bar{6}}}]^{i\ \, s}_{\ j \ \, t}$ \\[0.83ex]
 \hline
   \rule{0pt}{3.5ex}$\boldsymbol{3} \otimes \boldsymbol{6} \otimes \boldsymbol{6} \otimes \boldsymbol{\bar{6}}$ &  \multicolumn{3}{c|}{} & $\bt{J}^{\, s\,ia}[\bt{t}_{\boldsymbol{6}}^a]_u^{\ \, t}$ & & $\bt{Q}^{\, q\, i}_{\ \ \ u} \bt{W}_q^{\ \, st}$ & & $\ni [\Pi_{\boldsymbol{366\bar{6}}}]^{i\, st}_{\ \ \ \,u}$ \\[0.83ex]
 \hline
   \rule{0pt}{3.5ex}$\boldsymbol{3} \otimes \boldsymbol{6} \otimes \boldsymbol{8} \otimes \boldsymbol{8}$ & \ $[\bt{t}_{\boldsymbol{3}}^a]_j^{\ \, i} \bt{J}^{\, s\, jb}$ \ \ & & $\bt{J}^{\,t\,ia} [\bt{t}_{\boldsymbol{6}}^b]_t^{\ \, s}$ & $\bt{J}^{\,s\,ic}\{f, d\}^{abc}$ & $\bt{E}_x^{\ \,is}\bt{G}^{\,x\,ab}$ & $\bt{V}_q^{\ \, ia}\bt{X}^{q\, sb}$ & & $\ni [\Pi_{\boldsymbol{3688}}]^{i\,s\,ab}$\\[0.83ex]
 \hline
   \rule{0pt}{3.5ex}$\boldsymbol{3} \otimes \boldsymbol{\bar{6}} \otimes \boldsymbol{\bar{6}} \otimes \boldsymbol{8}$ &  $\bt{K}_s^{\ \,ij}\, \bar{\bt{J}}_{s\, ai}$ & & $\bt{J}^{\, s\,ia} \bar{\bt{S}}_{stu}$ & \multicolumn{4}{c|}{} & $\ni [\Pi_{\boldsymbol{3\bar{6}\bar{6}8}}]^{i\ \ \ a}_{\ st}$ \\[0.83ex]
 \hline
  \rule{0pt}{3.5ex}$\boldsymbol{6} \otimes \boldsymbol{6} \otimes \boldsymbol{\bar{6}} \otimes \boldsymbol{\bar{6}}$ &  $\delta_u^{\ s}\,\delta_v^{\ t}$ & \multicolumn{2}{c|}{} & $[\bt{t}_{\boldsymbol{6}}^a]_u^{\ \,s}[\bt{t}_{\boldsymbol{6}}^a]_v^{\ \,t}$ & \multicolumn{3}{c|}{} & $\ni [\Pi_{\boldsymbol{66\bar{6}\bar{6}}}]^{st}_{\ \ uv}$ \\[0.83ex]
 \hline
   \rule{0pt}{3.5ex}$\boldsymbol{6} \otimes \boldsymbol{6} \otimes \boldsymbol{6} \otimes \boldsymbol{8}$ & \multicolumn{3}{c|}{} & $\bt{S}^{\, str}[\bt{t}_{\boldsymbol{6}}^a]_r^{\ \,u}$ & & $\bt{W}_q^{\ \,st} \bt{X}^{q\, ua}$ & & $\ni [\Pi_{\boldsymbol{6668}}]^{stu\,a}$ \\[0.83ex]
 \hline
   \rule{0pt}{3.5ex}$\boldsymbol{6} \otimes \boldsymbol{\bar{6}} \otimes \boldsymbol{8} \otimes \boldsymbol{8}$ &  $\delta_t^{\ s}\, \delta_b^{\ a}$ & \multicolumn{2}{c|}{} & $[\bt{t}_{\boldsymbol{6}}^c]_t^{\ \, s}[\bt{t}_{\boldsymbol{8}}^c]_b^{\ \, a}$ &  \multicolumn{2}{c|}{} & \ $\bt{F}^{\,n\ s}_{\ \ t} \bt{H}^{\,n\,ab}$\ \ & \ $\ni [\Pi_{\boldsymbol{6\bar{6}88}}]^{s\ a}_{\ t\ \, b}$\ \ \\[0.83ex]
 \hline
 \end{tabular}\\[0.83ex]
 \end{center}
 \caption{Color index contractions yielding the four-field color invariants needed for the operator catalog, ordered by the dimension of the representation whose index is summed over.}
 \end{table*}
\renewcommand*{\arraystretch}{1}

\renewcommand*{\arraystretch}{2}
\begin{table*}[t!]\label{t3}
\begin{center}
 \begin{tabular}{|c||c|c|c|c|}
\hline
\rule{0pt}{3.4ex}Invariant & \multicolumn{3}{c|}{\ Clebsch-Gordan coefficients\ \ } & Notes \\[0.83ex]
\hline
\hline
   \rule{0pt}{3.5ex}\ $\boldsymbol{3} \otimes \boldsymbol{3} \otimes \boldsymbol{3} \otimes \boldsymbol{3} \otimes \boldsymbol{6}$\ \ & $[\Pi_{\boldsymbol{3366}}]^{ij\,st}\bt{K}_t^{\ \,kl}$ & \multicolumn{2}{c|}{} & \ $\ni [\Upsilon_{\boldsymbol{33336}}]^{ijkl\,s}$\ \ \\[0.83ex]
 \hline
   \rule{0pt}{3.5ex}\ $\boldsymbol{3} \otimes \boldsymbol{3} \otimes \boldsymbol{3} \otimes \boldsymbol{\bar{3}} \otimes \boldsymbol{\bar{6}}$\ \ & \ $[\Pi_{\boldsymbol{3\bar{3}6\bar{6}}}]^{i\ \, t}_{\ l \ \, s}\, \bt{K}_t^{\ \, jk}$\ \ & $[\Pi_{\boldsymbol{33\bar{6}8}}]^{ij\ \,a}_{\ \ s} [\bt{t}_{\boldsymbol{3}}^a]_l^{\ \,k}$ & & $\ni [\Upsilon_{\boldsymbol{333\bar{3}\bar{6}}}]^{ijk}_{\ \ \ \,l\,s}$ \\[0.83ex]
 \hline
   \rule{0pt}{3.5ex}\ $\boldsymbol{3} \otimes \boldsymbol{3} \otimes \boldsymbol{6} \otimes \boldsymbol{\bar{6}} \otimes \boldsymbol{\bar{6}}$\ \ & $[\Pi_{\boldsymbol{3366}}]^{ij\,sr}\bar{\bt{S}}_{rtu}$ & $[\Pi_{\boldsymbol{33\bar{6}8}}]^{ij\ \, a}_{\ \ t}[\bt{t}_{\boldsymbol{6}}^a]_u^{\ \, s}$ & & $\ni [\Upsilon_{\boldsymbol{336\bar{6}\bar{6}}}]^{ij\,s}_{\ \ \ \, tu}$ \\[0.83ex]
 \hline
   \rule{0pt}{3.5ex}\ $\boldsymbol{3} \otimes \boldsymbol{\bar{3}} \otimes \boldsymbol{6} \otimes \boldsymbol{6} \otimes \boldsymbol{6}$\ \ & $[\Pi_{\boldsymbol{3\bar{3}6\bar{6}}}]^{i\ \, s}_{\ j \ \, r}\, \bt{S}^{\,rtu}$ & $[\bar{\Pi}_{\boldsymbol{3\bar{6}\bar{6}8}}]_{j\ \ \, a}^{\ \, st}\, \bt{J}^{\,u\,ia}$ & \ $[\Pi_{\boldsymbol{6668}}]^{stu\,a}[\bt{t}_{\boldsymbol{3}}^a]_j^{\ \,i}$\ \ & \ $\ni [\Upsilon_{\boldsymbol{3\bar{3}666}}]^{i\ \, stu}_{\ j}$\ \ \\[0.83ex]
 \hline
   \rule{0pt}{3.5ex}\ $\boldsymbol{3} \otimes \boldsymbol{\bar{3}} \otimes \boldsymbol{6} \otimes \boldsymbol{\bar{6}} \otimes \boldsymbol{8}$\ \ & & $[\Pi_{\boldsymbol{3688}}]^{i\,s\,ab}\, \bar{\bt{J}}_{t\,bj}$ & $[\Pi_{\boldsymbol{6\bar{6}88}}]^{s\ a}_{\ t\ \, b}\, [\bt{t}_{\boldsymbol{3}}^b]_j^{\ \,i}$  & $\ni [\Upsilon_{\boldsymbol{3\bar{3}6\bar{6}8}}]^{i\ \,s\ \,a}_{\ j\ \,t}$ \\[0.83ex]
 \hline
   \rule{0pt}{3.5ex}\ $\boldsymbol{3} \otimes \boldsymbol{6} \otimes \boldsymbol{6} \otimes \boldsymbol{6} \otimes \boldsymbol{6}$\ \ & $[\Pi_{\boldsymbol{366\bar{6}}}]^{i\,st}_{\ \ \ r}\,\bt{S}^{\,ruv}$ & \multicolumn{2}{c|}{} & $\ni [\Upsilon_{\boldsymbol{36666}}]^{i\,stuv}$ \\[0.83ex]
 \hline
   \rule{0pt}{3.5ex}\ $\boldsymbol{3} \otimes \boldsymbol{6} \otimes \boldsymbol{\bar{6}} \otimes \boldsymbol{\bar{6}} \otimes \boldsymbol{\bar{6}}$\ \ & & $[\Pi_{\boldsymbol{3\bar{6}\bar{6}8}}]^{i\ \ \,a}_{\ tu}[\bt{t}_{\boldsymbol{6}}^a]_v^{\ \,s}$ & $[\bar{\Pi}_{\boldsymbol{6668}}]_{tuv\,a}\bt{J}^{\,s\,ia}$ & $\ni [\Upsilon_{\boldsymbol{36\bar{6}\bar{6}\bar{6}}}]^{i\,s}_{\ \ \, tuv}$ \\[0.83ex]
 \hline
   \rule{0pt}{3.5ex}\ $\boldsymbol{6} \otimes \boldsymbol{6} \otimes \boldsymbol{\bar{6}} \otimes \boldsymbol{\bar{6}} \otimes \boldsymbol{8}$\ \ & $[\Pi_{\boldsymbol{66\bar{6}\bar{6}}}]^{st}_{\ \ ur}[\bt{t}_{\boldsymbol{6}}^a]_v^{\ \,r}$ & $[\Pi_{\boldsymbol{6\bar{6}88}}]^{s\ a}_{\ u\ \, b}[\bt{t}_{\boldsymbol{6}}^b]_v^{\ \,t}$ & & $\ni [\Upsilon_{\boldsymbol{66\bar{6}\bar{6}8}}]^{st\ \ \ a}_{\ \ uv}$ \\[0.83ex]
 \hline
   \rule{0pt}{3.5ex}\ $\boldsymbol{6} \otimes \boldsymbol{6} \otimes \boldsymbol{6} \otimes \boldsymbol{8} \otimes \boldsymbol{8}$\ \ & & \ $[\Pi_{\boldsymbol{6668}}]^{stu\,c}\{f,d\}^{abc}$\ \  & & $\ni [\Upsilon_{\boldsymbol{66688}}]^{stu\,ab}$ \\[0.83ex]
 \hline
   \rule{0pt}{3.5ex}\ $\boldsymbol{6} \otimes \boldsymbol{\bar{6}} \otimes \boldsymbol{\bar{6}} \otimes \boldsymbol{\bar{6}} \otimes \boldsymbol{\bar{6}}$\ \ & $[\Pi_{\boldsymbol{66\bar{6}\bar{6}}}]^{rs'}_{\ \ \ st}\,\bar{\bt{S}}_{s'uv}$ & \multicolumn{2}{c|}{} & $\ni [\Upsilon_{\boldsymbol{6\bar{6}\bar{6}\bar{6}\bar{6}}}]^{r}_{\ \,stuv}$ \\[0.83ex]
 \hline
  \end{tabular}\\[0.83ex]
 \end{center}
 \caption{Color index contractions yielding the required five-field color invariants.}
 \end{table*}
\renewcommand*{\arraystretch}{1}

Tables \hyperref[t1]{I}--\hyperref[t3]{III} display a wide variety of index contractions leading to color singlets. The number of valid contractions naturally rises with the order of the invariant, since higher-order contractions are built from smaller ones. Therefore, while most three-field invariants permit only one contraction, the five-field invariant $\boldsymbol{3} \otimes \boldsymbol{\bar{3}} \otimes \boldsymbol{6} \otimes \boldsymbol{6} \otimes \boldsymbol{6}$ (for example) permits six. There are in total a few dozen contractions available for all possible sets of color-charged fields relevant to our operator catalog, even before considering sextet spin(s) and electroweak representation(s) --- and the other fields required to form singlets under the full SM gauge group.

\subsection{Lorentz invariance}
\label{s2.2}

The construction of Lorentz-invariant operators naturally depends on the spin(s) of the exotic field(s).  We consider scenarios with one or two (distinct) color-sextet fields
assigned as either a complex scalar or a Dirac fermion (or one of each species). In order to collect all sextet operators, we identify all possible Lorentz structures built out of the $\mathrm{SU}(3)_{\text{c}}$ invariants collected above. Since it turns out that most operators in our catalog contain Dirac fermions --- either quarks, leptons, or color-sextet fermions --- a natural starting place is the non-vanishing Dirac bilinears, displayed in \hyperref[t4]{Table IV}, with their Hermitian conjugates implied.
\renewcommand*{\arraystretch}{2}
\begin{table*}[t]\label{t4}
\begin{center}
 \begin{tabular}{|c||c|c|c|}
\hline
\rule{0pt}{3.4ex} & Examples & Bilinears & Notes \\[0.5ex]
\hline
\hline
\multirow{3}{*}[-1.75ex]{\ $(\bar{\chi}\chi')$} \rule{0pt}{3.5ex} & \multirow{3}{*}[-1.75ex]{\shortstack{$(\bar{q}q')$ ($\boldsymbol{\bar{3}}\otimes \boldsymbol{3}$),\\ $(\bar{\Psi}\Psi)$ ($\boldsymbol{\bar{6}}\otimes \boldsymbol{6}$),\\ $(\bar{q}\ell)$ ($\boldsymbol{\bar{3}} \otimes \boldsymbol{1}$)}} & $\bar{X}_{\text{L}} H \Gamma \chi'_{\text{R}}$ & \\[0.83ex]
\cline{3-4}
 \rule{0pt}{3.5ex} &  & $\bar{X}_{\text{L}} \Omega\, \gamma^{\mu} X'_{\text{L}}$ & \multirow{2}{*}[-1ex]{\shortstack{only if half of\\ four-fermion operator\\ with second $\gamma_{\mu}$}} \\[0.83ex]
\cline{3-3}
 \rule{0pt}{3.5ex} & & $\bar{\chi}_{\text{R}}\gamma^{\mu}\chi_{\text{R}}'$ & \\[0.83ex]
 \hline
 \hline
   \multirow{3}{*}[-1.75ex]{\ $(\chi\chi')$} \rule{0pt}{3.5ex} & \multirow{3}{*}[-1.75ex]{\shortstack{$(qq')$ ($\boldsymbol{3} \otimes \boldsymbol{3})$,\\ $(\Psi\Psi)$ ($\boldsymbol{6} \otimes \boldsymbol{6}$),\\ $(q\ell)$ ($\boldsymbol{3} \otimes \boldsymbol{1}$)}} & $\overbar{\chi^{\text{c}}_{\text{R}}}\,\Gamma \chi'_{\text{R}}$ & \multirow{2}{*}[-1ex]{\shortstack{\ $\Gamma = \sigma^{\mu\nu}$ non-vanishing\ \ \\ only if $\chi' \neq \chi$}} \\[0.83ex]
\cline{3-3}
 \rule{0pt}{3.5ex} & & $\overbar{X^{\text{c}}_{\text{L}}} \,\Omega\, \Gamma X_{\text{L}}'$ & \\[0.83ex]
\cline{3-4}
 \rule{0pt}{3.5ex} & & $\overbar{X^{\text{c}}_{\text{L}}}\, H \gamma^{\mu} \chi'_{\text{R}}$ & needs second $\gamma_{\mu}$ again \\[0.83ex]
 \hline
\bottomrule
\multicolumn{2}{|c|}{Operator} & \multicolumn{2}{c|}{Notes} \\[0.5ex]
\hline
\hline
\multicolumn{2}{|c|}{\rule{0pt}{3.5ex}$\Gamma \in \{\boldsymbol{1}, \sigma^{\mu\nu}\}$} & \multicolumn{2}{c|}{\multirow{2}{*}[-1ex]{\shortstack{$\sigma^{\mu\nu}$ must be accompanied by\\ $\sigma_{\mu\nu}$ or a field-strength tensor $F_{\mu\nu}$,\\ $F_{\mu\nu} \in \{B_{\mu\nu}, \bt{t}_{\boldsymbol{2}}^A W^A_{\mu\nu}, G^a_{\mu\nu}\}$}}} \\[0.83ex]
\cline{1-2}
\multicolumn{2}{|c|}{\rule{0pt}{3.5ex}$\Omega \in \{H H^{\dagger}, \ii \tau^2\}$} & \multicolumn{2}{c|}{}\\[0.83ex]
\hline
 \end{tabular}\\[0.83ex]
 \end{center}
 \caption{Fermion bilinears that can be used, possibly by themselves or paired with an appropriate second bilinear, to build all Lorentz-invariant operators of mass dimension up to seven including at least one color-sextet field. Hermitian conjugates are allowed where distinct. Stipulations exist on the use of some bilinears listed above; some comments are offered where appropriate. Extra insertions of $|H|^2$ may be allowed in dimension-five operators where gauge invariant. The dual Higgs field $\tilde{H} = \ii \tau^2 H^{\dagger}$ could be used in place of $H$ anywhere to \emph{e.g.} allow a specific sextet weak hypercharge. CP-odd bilinears replacing $\boldsymbol{1}$ with $\gamma^5$ are also allowed in principle.}
 \end{table*}
\renewcommand*{\arraystretch}{1}

This table establishes a shorthand for non-vanishing Dirac bilinears in addition to cataloging these objects. The top block contains bilinears reminiscent of Standard Model structures, including all contractions of some generic weak doublet $X_{\text{L}} = (\chi^+_{\text{L}}, \chi^-_{\text{L}})^{\transpose}$, and a weak singlet $\chi_{\text{R}}$. This class of bilinears is useful for invariants containing direct products of the form $\boldsymbol{\bar{\textbf{r}}} \otimes \textbf{q}$ --- an irreducible representation with a (possibly distinct) conjugate irreducible representation. A familiar example is the SM operator
\begin{align}
    \mathcal{L}_{\text{SM}} \supset -y^d_{IJ}(\bar{Q}^i_{\text{L}I} H d_{\text{R}Ji}),
\end{align}
with $I,J \in \{1,2,3\}$ generation (flavor) indices, the coefficients of which form the down-type Yukawa matrix $y_{IJ}^d$. The color invariant underpinning this operator is $\boldsymbol{\bar{3}}\otimes \boldsymbol{3}$. We have found many higher-order color invariants that include a structure like this --- either between two color-charged fields or between a color-charged field and a lepton --- so these Dirac bilinears are quite useful. But we also need bilinears for direct products of the form $\textbf{r} \otimes \textbf{q}$ --- an irreducible representation with a (possibly distinct) irreducible representation. These bilinears are displayed in the middle block of \hyperref[t4]{Table IV}. They comprise Dirac fermions and charge-conjugated fermions:
\begin{align}\label{CC}
\chi^{\text{c}} \equiv \text{C}\bar{\chi}^{\transpose}\ \ \ \text{with C satisfying}\ \ \ \text{C}\gamma^{\mu} \text{C}^{-1} = -\gamma^{\mu}.
\end{align}
While these structures do not appear in the Standard Model, some occur in a variety of bSM theories ranging from models of color-sextet scalars \cite{Han_2010_2} to supersymmetric models with Dirac gauginos \cite{Carpenter:2020mrsm}. The lower block of \hyperref[t4]{Table IV} displays various objects in weak or Dirac space that can be inserted in the bilinears listed above. Omitted from this particular list is the fifth Dirac matrix $\gamma^5$, which when inserted between two Dirac fields of definite chirality, as considered in this work, modifies the bilinear only by a global sign. Crucially, some of these objects are themselves Standard Model fields, including any of the SM field-strength tensors\footnote{The $\mathrm{SU}(2)_{\text{L}}$ (weak-isospin) field-strength tensor $W^{\mu\nu}$ must be accompanied by $\bt{t}_{\boldsymbol{2}}^A = \tau^A/2$, $A \in \{1,2,3\}$, the generators of the fundamental representation $\boldsymbol{2}$ of $\mathrm{SU}(2)$. These generators must furthermore be contracted with $\mathrm{SU}(2)_{\text{L}}$ doublets $X_{\text{L}}$.} $B^{\mu\nu},W^{\mu\nu},G^{\mu\nu}$ and the SM Higgs doublet, which in the unitary gauge takes the form
\begin{align}
    H = \frac{1}{\sqrt{2}}\begin{pmatrix}
    0\\
    v+h\end{pmatrix}.
\end{align}
As is well known, operators containing insertions of the Higgs doublet generate operators of lower ``effective dimension'' when the Higgs is replaced by its vacuum expectation value $v$. A complete operator catalog follows from cycling through all of the bilinears and insertions that satisfy our criteria for gauge and Lorentz invariance.

\renewcommand*{\arraystretch}{2}
\begin{table*}[t]\label{t5}
\begin{center}
 \begin{tabular}{|c||c|c||c|c||c|c|}
\hline
\rule{0pt}{3.4ex} & \multicolumn{2}{c||}{\ \emph{Scalar} sextet $\Phi$ only\ \ } & \multicolumn{2}{c||}{\ \emph{Dirac} sextet $\Psi$ only\ \ } & \multicolumn{2}{c|}{$\geq 1$ of each} \\[0.5ex]
\hline
\rule{0pt}{3.4ex}\ $\mathrm{SU}(3)_{\text{c}}$ invariant\ \ & \ $d_{\text{min}}$\ \ & Structure & \ $d_{\text{min}}$\ \ & Structure & \ $d_{\text{min}}$\ \ & \ Structure\ \ \\[0.83ex]
\hline
\hline
\multirow{3}{*}[-1.75ex]{\ $\boldsymbol{6} \otimes \boldsymbol{\bar{6}}$} \rule{0pt}{3.5ex} & $4^{\dagger}$ & $\Phi^{\dagger}\Phi$ & $5^{\dagger}$ & $(\bar{\Psi}\Psi)$ & \multirow{3}{*}[-1.75ex]{$4$} & $(\bar{\Psi}\ell)\Phi$ \\[0.83ex]
\cline{2-5}
\cline{7-7}
 \rule{0pt}{3.5ex} & \multicolumn{2}{c||}{\multirow{2}{*}[0ex]{}} & \multicolumn{2}{c||}{} & & $(\Psi \ell)\Phi^{\dagger}$ \\[0.83ex]
 \cline{7-7}
 \rule{0pt}{3.5ex} & \multicolumn{2}{c||}{} & \multicolumn{2}{c||}{} &  & $(\bar{\ell}\Psi)\Phi^{\dagger}$ \\[0.83ex]
 \hline
 \hline
\multirow{3}{*}[-1.75ex]{\ $\boldsymbol{3} \otimes \boldsymbol{3} \otimes \boldsymbol{\bar{6}}$} \rule{0pt}{3.5ex} & 4 & $(qq')\Phi^{\dagger}$ & \multirow{3}{*}[-1.75ex]{6} & $(qq')(\bar{\Psi}\ell)$ & \multicolumn{2}{c|}{}\\[0.83ex]
\cline{2-3}
\cline{5-5}
 \rule{0pt}{3.5ex} & 6 & $(qq')|H|^2\Phi^{\dagger}$ & & $(\bar{\Psi}q)(q\ell)$ & \multicolumn{2}{c|}{} \\[0.83ex]
 \cline{2-3}
\cline{5-5}
 \rule{0pt}{3.5ex} & \multicolumn{2}{c||}{} & & $(\bar{\Psi}q)(\bar{\ell}q)$ & \multicolumn{2}{c|}{} \\[0.83ex]
 \hline
 \hline
\multirow{2}{*}[-1ex]{\ $\boldsymbol{3} \otimes \boldsymbol{6} \otimes \boldsymbol{8}$} \rule{0pt}{3.5ex} & \multirow{2}{*}[-1ex]{6} & $(q\ell)\Phi G$ & 5 & $(q\Psi)G$ & \multicolumn{2}{c|}{}\\[0.83ex]
\cline{3-3}
\cline{4-5}
 \rule{0pt}{3.5ex} & & $(\bar{\ell}q)\Phi G$ & 7 & $(q\Psi)|H|^2 G$ & \multicolumn{2}{c|}{} \\[0.83ex]
 \hline
 \hline
 \multirow{3}{*}[-1.75ex]{\ $\boldsymbol{6} \otimes \boldsymbol{6} \otimes \boldsymbol{6}$} \rule{0pt}{3.5ex} & $5^{\dagger}$ & $\Phi\Phi\Phi$ & \multirow{2}{*}[-1ex]{6} & $(\Psi\Psi)(\Psi\ell)$ & \multirow{2}{*}[-1ex]{5} & $(\Psi \ell)\Phi\Phi$\\[0.83ex]
\cline{2-3}
\cline{5-5}
\cline{7-7}
 \rule{0pt}{3.5ex} & \multicolumn{2}{c||}{} & & $(\Psi\Psi)(\bar{\ell}\Psi)$ & & $(\bar{\ell}\Psi)\Phi\Phi$ \\[0.83ex]
 \cline{4-5}
 \cline{6-7}
  \rule{0pt}{3.5ex} & \multicolumn{2}{c||}{} &  \multicolumn{2}{c||}{} & $6^{\dagger}$ & $(\Psi\Psi)\Phi$ \\[0.83ex]
 \hline
 \hline
 \multirow{3}{*}[-1.75ex]{\ $\boldsymbol{6} \otimes \boldsymbol{\bar{6}} \otimes \boldsymbol{8}$} \rule{0pt}{3.5ex} & 6 & $\Phi^{\dagger}\Phi GB$ & 5 & $(\bar{\Psi}\Psi)G$ & \multirow{3}{*}[-1.75ex]{6} & $(\bar{\Psi}\ell)\Phi G$\\[0.83ex]
\cline{2-5}
\cline{7-7}
 \rule{0pt}{3.5ex} & \multicolumn{2}{c||}{} & 7 & $(\bar{\Psi}\Psi)|H|^2G$ & & $(\Psi \ell)\Phi^{\dagger} G$ \\[0.83ex]
 \cline{4-5}
 \cline{7-7}
  \rule{0pt}{3.5ex} & \multicolumn{2}{c||}{} & \multicolumn{2}{c||}{} & & $(\bar{\ell}\Psi)\Phi^{\dagger} G$ \\[0.83ex]
\hline
 \end{tabular}\\[0.83ex]
 \end{center}
 \caption{Schematic table of three-field invariant operators, plus the unique two-field invariant $\boldsymbol{6} \otimes \boldsymbol{\bar{6}}$, of minimum mass dimension $d_{\text{min}} \leq 7$ that can be constructed using the fermion bilinears in \hyperref[t4]{Table IV} and that include at least one color-sextet field. We consider scenarios with a sextet scalar, a sextet (Dirac) fermion, and at least one of each species. Note that operators requiring a single gluon field-strength tensor $G$ must be made Lorentz invariant by judicious choice of fermion bilinear(s) or a weak-hypercharge field strength $B$. Lists marked with ${}^\dagger$ have indicated $d_{\text{min}}$ once accompanied by minimal set of SM fields.}
 \end{table*}
\renewcommand*{\arraystretch}{1}
\renewcommand*{\arraystretch}{2}
\begin{table*}[ht!]\label{t6}
\begin{center}
 \begin{tabular}{|c||c|c||c|c||c|c|}
\hline
\rule{0pt}{3.4ex} & \multicolumn{2}{c||}{\ \emph{Scalar} sextet $\Phi$ only\ \ } & \multicolumn{2}{c||}{\ \emph{Dirac} sextet $\Psi$ only\ \ } & \multicolumn{2}{c|}{$\geq 1$ of each} \\[0.5ex]
\hline
\rule{0pt}{3.4ex}\ $\mathrm{SU}(3)_{\text{c}}$ invariant\ \ & \ $d_{\text{min}}$\ \ & Structure & \ $d_{\text{min}}$\ \ & Structure & \ $d_{\text{min}}$\ \ & Structure\\[0.83ex]
\hline
\hline
\multirow{2}{*}[-1ex]{\ $\boldsymbol{3} \otimes \boldsymbol{3} \otimes \boldsymbol{6} \otimes \boldsymbol{6}$} \rule{0pt}{3.5ex} & 5 & $(qq')\Phi \Phi$ & \multirow{2}{*}[-1ex]{6} & $(qq')(\Psi\Psi)$ & \multirow{2}{*}[-1ex]{7} & $(qq')(\Psi\ell)\Phi$\\[0.83ex]
\cline{2-3}
\cline{5-5}
\cline{7-7}
 \rule{0pt}{3.5ex} & 7 & $(qq')\Phi |H|^2 \Phi$ & & $(q\Psi)(q'\Psi)$ & & $(q\ell)(q'\Psi)\Phi$ \\[0.83ex]
 \hline
 \hline
 \ $\boldsymbol{3} \otimes \boldsymbol{3} \otimes \boldsymbol{\bar{6}} \otimes \boldsymbol{8}$ \rule{0pt}{3.5ex} & 6 & $(qq')\Phi^{\dagger} G$ & \multicolumn{2}{c||}{} &  \multicolumn{2}{c|}{} \\[0.83ex]
\hline
\hline
\multirow{2}{*}[-1ex]{\ $\boldsymbol{3} \otimes \boldsymbol{\bar{3}} \otimes \boldsymbol{\bar{3}} \otimes \boldsymbol{\bar{6}}$} \rule{0pt}{3.5ex} & \multirow{2}{*}[-1ex]{7} & $(\bar{q}q')(\bar{q}''\ell)\Phi$ & \multirow{2}{*}[-1ex]{6} & $(\bar{q}q')(\bar{q}''\Psi)$ & \multicolumn{2}{c|}{\multirow{2}{*}[-1ex]{}}\\[0.83ex]
\cline{3-3}
\cline{5-5}
\rule{0pt}{3.5ex} & & $(qq')^{\dagger}(q''\ell)\Phi$ & & $(qq')^{\dagger}(\bar{q}''\Psi)$ & \multicolumn{2}{c|}{} \\[0.83ex]
\hline
\hline
\multirow{2}{*}[-1ex]{\ $\boldsymbol{3} \otimes \boldsymbol{\bar{3}} \otimes \boldsymbol{6} \otimes \boldsymbol{\bar{6}}$} \rule{0pt}{3.5ex} & 5 & $(\bar{q}q')\Phi^{\dagger} \Phi$ & \multirow{2}{*}[-1ex]{6} & $(\bar{q}q')(\bar{\Psi}\Psi)$ & \multirow{2}{*}[-1ex]{7$^*$} & $(\bar{q}q')(\bar{\Psi}\ell)\Phi$\\[0.83ex]
\cline{2-3}
\cline{5-5}
\cline{7-7}
 \rule{0pt}{3.5ex} & 7 & $(\bar{q}q')\Phi^{\dagger} |H|^2 \Phi$ & & $(\bar{q}\Psi)(\bar{\Psi}q')$ & & $(\bar{q}\Psi)(q'\ell)\Phi^{\dagger}$ \\[0.83ex]
 \hline
 \hline
\multirow{4}{*}[-2ex]{\ $\boldsymbol{3} \otimes \boldsymbol{6} \otimes \boldsymbol{6} \otimes \boldsymbol{\bar{6}}$} \rule{0pt}{3.5ex} & \multirow{2}{*}[-1ex]{6} & $(q\ell)|\Phi|^2 \Phi$ & \multirow{2}{*}[-1ex]{6} & $(q\Psi)(\bar{\Psi}\Psi)$ & \multirow{2}{*}[-1ex]{5} & $(q\Psi)\Phi^{\dagger}\Phi$ \\[0.83ex]
\cline{3-3}
\cline{5-5}
\cline{7-7}
 \rule{0pt}{3.5ex} &  & $(\bar{\ell}q)|\Phi|^2\Phi$ & & $(\bar{\Psi}q)(\Psi\Psi)$ & & $(\bar{\Psi}q)\Phi \Phi$ \\[0.83ex]
 \cline{2-7}
  \rule{0pt}{3.5ex} & \multicolumn{2}{c||}{} & \multicolumn{2}{c||}{} & \multirow{2}{*}[-1ex]{7$^*$} & $(q\Psi)(\Psi\ell)\Phi^{\dagger}$ \\[0.83ex]
  \cline{7-7}
    \rule{0pt}{3.5ex} & \multicolumn{2}{c||}{} & \multicolumn{2}{c||}{} & & $(q\ell)(\bar{\Psi}\Psi)\Phi$ \\[0.83ex]
 \hline
 \hline
  \ $\boldsymbol{3} \otimes \boldsymbol{6} \otimes \boldsymbol{8} \otimes \boldsymbol{8}$ \rule{0pt}{3.5ex} & \multicolumn{2}{c||}{} & 7 & $(q\Psi)GG$ & \multicolumn{2}{c|}{}\\[0.83ex]
\hline
\hline
\multirow{2}{*}[-1ex]{\ $\boldsymbol{3} \otimes \boldsymbol{\bar{6}} \otimes \boldsymbol{\bar{6}} \otimes \boldsymbol{8}$} \rule{0pt}{3.5ex} & \multirow{2}{*}[-1ex]{7} & $(q\ell)\Phi^{\dagger}\Phi^{\dagger}G$ & \multicolumn{2}{c||}{} & 6 & $(\bar{\Psi}q)\Phi^{\dagger} G$ \\[0.83ex]
\cline{3-3}
\cline{6-7}
 \rule{0pt}{3.5ex} &  & $(\bar{\ell}q)\Phi^{\dagger}\Phi^{\dagger} G$ & \multicolumn{2}{c||}{} & \multicolumn{2}{c|}{} \\[0.83ex]
 \hline
 \hline
  \multirow{4}{*}[-2ex]{\ $\boldsymbol{6} \otimes \boldsymbol{6} \otimes \boldsymbol{\bar{6}} \otimes \boldsymbol{\bar{6}}$} \rule{0pt}{3.5ex} & $6^{\dagger}$ & $|\Phi|^4$ & \multicolumn{2}{c||}{} & \multirow{3}{*}[-1.75ex]{6} & $(\bar{\Psi}\ell)|\Phi|^2\Phi$\\[0.83ex]
   \cline{2-3}
   \cline{7-7}
   \rule{0pt}{3.5ex} & \multicolumn{2}{c||}{} & \multicolumn{2}{c||}{} &  & $(\Psi\ell)|\Phi|^2\Phi^{\dagger}$\\[0.83ex]
   \cline{7-7}
 \rule{0pt}{3.5ex} & \multicolumn{2}{c||}{} & \multicolumn{2}{c||}{} &  & $(\bar{\ell}\Psi)|\Phi|^2\Phi^{\dagger}$\\[0.83ex]
 \cline{6-7}
  \rule{0pt}{3.5ex} & \multicolumn{2}{c||}{} & \multicolumn{2}{c||}{} & 7 & \ $(\bar{\Psi}\Psi)|\Phi|^2|H|^2$\ \ \\[0.83ex]
\hline
\hline
 \multirow{3}{*}[-1.75ex]{\ $\boldsymbol{6} \otimes \boldsymbol{6} \otimes \boldsymbol{6} \otimes \boldsymbol{8}$} \rule{0pt}{3.5ex} & 7 & $\Phi \Phi \Phi G B$ & \multicolumn{2}{c||}{} & 6 & $(\Psi \Psi)\Phi G$\\[0.83ex]
   \cline{2-3}
   \cline{6-7}
   \rule{0pt}{3.5ex} & \multicolumn{2}{c||}{} & \multicolumn{2}{c||}{} & \multirow{2}{*}[-1ex]{7} & $(\Psi\ell)\Phi\Phi G$\\[0.83ex]
   \cline{7-7}
      \rule{0pt}{3.5ex} & \multicolumn{2}{c||}{} & \multicolumn{2}{c||}{} &  & $(\bar{\ell}\Psi)\Phi\Phi G$\\[0.83ex]
\hline
\hline
   \ $\boldsymbol{6} \otimes \boldsymbol{\bar{6}} \otimes \boldsymbol{8} \otimes \boldsymbol{8}$ \rule{0pt}{3.5ex} & 6 & $|\Phi|^2 G G$ & 7 & $(\bar{\Psi}\Psi) GG$ & \multicolumn{2}{c|}{}\\[0.83ex]
\hline
 \end{tabular}\\[0.83ex]
 \end{center}
 \caption{Schematic table of four-field invariant operators. Fields are left blank if operators exist only with $d_{\text{min}} = 8$. Lists marked with $^*$ are not exhaustive. Lists marked with ${}^{\dagger}$ have indicated $d_{\text{min}}$ once accompanied by minimal set of SM fields.}
 \end{table*}
\renewcommand*{\arraystretch}{1}
\renewcommand*{\arraystretch}{2}
\begin{table*}[t]\label{t7}
\begin{center}
 \begin{tabular}{|c||c|c||c|c|}
\hline
\rule{0pt}{3.4ex} & \multicolumn{2}{c||}{\ \emph{Scalar} sextet $\Phi$ only\ \ } & \multicolumn{2}{c|}{$\geq 1$ of each} \\[0.5ex]
\hline
\rule{0pt}{3.4ex}\ $\mathrm{SU}(3)_{\text{c}}$ invariant\ \ & \ $d_{\text{min}}$\ \ & Structure & \ $d_{\text{min}}$\ \ & Structure\\[0.83ex]
\hline
\hline
\rule{0pt}{3.5ex}\ $\boldsymbol{3} \otimes \boldsymbol{3} \otimes \boldsymbol{3} \otimes \boldsymbol{3} \otimes \boldsymbol{6}$\ \ & 7 & $(qq')(q''q''') \Phi$ & \multicolumn{2}{c|}{}\\[0.83ex]
\hline
\ $\boldsymbol{3} \otimes \boldsymbol{3} \otimes \boldsymbol{3} \otimes \boldsymbol{\bar{3}} \otimes \boldsymbol{\bar{6}}$ \rule{0pt}{3.5ex} & 7 & $(qq')(\bar{q}''q''')\Phi^{\dagger}$ & \multicolumn{2}{c|}{} \\[0.83ex]
\hline
\multirow{2}{*}[-1ex]{\ $\boldsymbol{3} \otimes \boldsymbol{3} \otimes \boldsymbol{6} \otimes \boldsymbol{\bar{6}} \otimes \boldsymbol{\bar{6}}$} \rule{0pt}{3.5ex} & 6 & $(qq')|\Phi|^2\Phi^{\dagger}$ & \multirow{2}{*}[-1ex]{$7^*$} & $(qq')(\bar{\Psi}\Psi)\Phi^{\dagger}$\\[0.83ex]
\cline{2-3}
\cline{5-5}
\rule{0pt}{3.5ex} & \multicolumn{2}{c||}{} & & $(\bar{\Psi}q)(q'\Psi)\Phi^{\dagger}$ \\[0.83ex]
\hline
\multirow{2}{*}[-1ex]{\ $\boldsymbol{3} \otimes \boldsymbol{\bar{3}} \otimes \boldsymbol{6} \otimes \boldsymbol{6} \otimes \boldsymbol{6}$} \rule{0pt}{3.5ex} & 6 & $(\bar{q}q')\Phi\Phi\Phi$ & \multirow{2}{*}[-1ex]{7} & $(\bar{q}q')(\Psi\Psi)\Phi$\\[0.83ex]
\cline{2-3}
\cline{5-5}
 \rule{0pt}{3.5ex} & \multicolumn{2}{c||}{} & & $(\bar{q}\Psi)(q'\Psi)\Phi$\\[0.83ex]
 \hline
\ $\boldsymbol{3} \otimes \boldsymbol{\bar{3}} \otimes \boldsymbol{6} \otimes \boldsymbol{\bar{6}} \otimes \boldsymbol{8}$ \rule{0pt}{3.5ex} & 7 & $(\bar{q}q')|\Phi|^2 G$ & \multicolumn{2}{c|}{} \\[0.83ex]
 \hline
\multirow{2}{*}[-1ex]{\ $\boldsymbol{3} \otimes \boldsymbol{6} \otimes \boldsymbol{6} \otimes \boldsymbol{6} \otimes \boldsymbol{6}$} \rule{0pt}{3.5ex} & \multirow{2}{*}[-1ex]{7} & $(q\ell)\Phi\Phi\Phi\Phi$ & \multicolumn{2}{c|}{\multirow{2}{*}[-1ex]{}}\\[0.83ex]
\cline{3-3}
\rule{0pt}{3.5ex} & & $(\bar{\ell}q)\Phi\Phi\Phi\Phi$ & \multicolumn{2}{c|}{} \\[0.83ex]
\hline
\multirow{2}{*}[-1ex]{\ $\boldsymbol{3} \otimes \boldsymbol{6} \otimes \boldsymbol{\bar{6}} \otimes \boldsymbol{\bar{6}} \otimes \boldsymbol{\bar{6}}$} \rule{0pt}{3.5ex} & \multirow{2}{*}[-1ex]{7} & $(q\ell)|\Phi|^2 \Phi^{\dagger}\Phi^{\dagger}$ & \multicolumn{2}{c|}{\multirow{2}{*}[-1ex]{}} \\[0.83ex]
\cline{3-3}
 \rule{0pt}{3.5ex} &  & $(\bar{\ell}q)|\Phi|^2\Phi^{\dagger}\Phi^{\dagger}$ & \multicolumn{2}{c|}{} \\[0.83ex]
 \hline
\ $\boldsymbol{6} \otimes \boldsymbol{6} \otimes \boldsymbol{\bar{6}} \otimes \boldsymbol{\bar{6}} \otimes \boldsymbol{8}$ \rule{0pt}{3.5ex} & \multicolumn{2}{c||}{} & 7 & $(\bar{\Psi} \Psi)\Phi^{\dagger}\Phi G$\\[0.83ex]
\hline
   \ $\boldsymbol{6} \otimes \boldsymbol{6} \otimes \boldsymbol{6} \otimes \boldsymbol{8} \otimes \boldsymbol{8}$ \rule{0pt}{3.5ex} & 7 & $\Phi\Phi\Phi G G$ & \multicolumn{2}{c|}{}\\[0.83ex]
\hline
   \ $\boldsymbol{6} \otimes \boldsymbol{\bar{6}} \otimes \boldsymbol{\bar{6}} \otimes \boldsymbol{\bar{6}} \otimes \boldsymbol{\bar{6}}$ \rule{0pt}{3.5ex} & $7$ & \ $\Phi\Phi^{\dagger}\Phi^{\dagger}\Phi^{\dagger}\Phi^{\dagger}|H|^2$\ \ & \multicolumn{2}{c|}{}\\[0.83ex]
\hline
 \end{tabular}\\[0.83ex]
 \end{center}
 \caption{Schematic table of five-field invariant operators. Here we consider scenarios with a sextet scalar and at least one of each species, since for these invariants there are no suitable fermion-only operators. Lists marked with $^*$ are not exhaustive.}
 \end{table*}
\renewcommand*{\arraystretch}{1}

The desired catalog of operators governing color-sextet fields interacting with the Standard Model can be constructed from the set of Dirac bilinears and color invariants.
The number of dimension-five and -six operators containing at least one sextet and at least one SM field is of $\mathcal{O}(10^2)$, and thus is quite large. Consequently,
we limit the listings in Tables \hyperref[t5]{V}--\hyperref[t7]{VII} to the \emph{minimal field content} at each order in the EFT expansion for each color invariant. These ``schematic'' operators are built using the shorthand Dirac bilinears introduced in \hyperref[t4]{Table IV}, with the generic Dirac fermions $\chi,\chi'$ in that table replaced by quarks, leptons, and weak singlet color-sextet fermions. Here we use a similar shorthand notation in which the familiar SM quarks and leptons with quantum numbers $(\mathrm{SU}(3)_{\text{c}},\mathrm{SU}(2)_{\text{L}},\mathrm{U}(1)_Y)$ are denoted by
\begin{align}\label{glossary}
q \in \begin{cases} Q_{\text{L}I} \sim (\boldsymbol{3},\boldsymbol{2},\tfrac{1}{6})\\[1.3ex]
\,u_{\text{R}I} \sim (\boldsymbol{3},\boldsymbol{1},\tfrac{2}{3})\\[1.3ex]
\,d_{\text{R}I} \sim (\boldsymbol{3},\boldsymbol{1},-\tfrac{1}{3})
\end{cases}\!\!\! \text{and}\ \ \ \ell \in \begin{cases} L_{\text{L}X} \sim (\boldsymbol{1},\boldsymbol{2},-\tfrac{1}{2})\\[1.3ex]
\,\ell_{\text{R}X} \sim (\boldsymbol{1},\boldsymbol{1},-1),
\end{cases}
\end{align}
where $I,X\in\{1,2,3\}$ are generation indices; and in which
\begin{align}\label{sextet_numbers}
\Phi \sim (\boldsymbol{6},\boldsymbol{1},Y_{\Phi})\ \ \ \text{and}\ \ \ \Psi \sim (\boldsymbol{6},\boldsymbol{1},Y_{\Psi})
\end{align}
respectively denote a color-sextet scalar and fermion. The gluon and weak-hypercharge field strength tensors appear explicitly as $G$ and $B$, as does the SM Higgs doublet $H$. Each entry in Tables \hyperref[t5]{V}--\hyperref[t7]{VII}, therefore, represents a set of operators given by appropriately combining all valid Lorentz structures with all available color index contractions as displayed in Tables \hyperref[t1]{I}--\hyperref[t3]{III}. $\mathrm{SU}(2)_{\text{L}}$ invariance generally has to be ensured by judicious choices of quark/lepton bilinears, but it is straightforward to preserve $\mathrm{U}(1)_Y$ by fixing the sextet hypercharge(s) after all of the other ingredients are specified.  Most operators can be generalized beyond the minimal field content (to even higher-dimensional operators) by insertions of Higgs or gauge boson in their Dirac bilinears.  These tables represent a comprehensive catalog at dimensions five and six, and also include a variety of interesting dimension-seven operators as well.

\subsection{Examples}

It is clear upon inspection of Tables \hyperref[t5]{V}--\hyperref[t7]{VII} that this catalog contains a formidable variety of gauge-invariant interactions for color-sextet fields. Not only do we recoup the fairly small set of interactions that have previously been investigated between sextet scalars and quark pairs $qq'$ \cite{Shu:2009xf,Han_2010,Han_2010_2} --- but we find many sextet interactions with quarks and a lepton, and many of these permit gauge bosons pursuant to either a color invariant or a Dirac bilinear. The operators become increasingly spectacular for the higher-order invariants: even at dimension six, for instance, there are triple-sextet interactions with quark pairs in \hyperref[t7]{Table VII}. In order to demonstrate how it can be used to develop a concrete model, we expand two subsets of the catalog and provide the associated operators explicitly. The sections we expand correspond to the three-field invariants $\boldsymbol{3} \otimes \boldsymbol{3} \otimes \boldsymbol{\bar{6}}$ and $\boldsymbol{3} \otimes \boldsymbol{6} \otimes \boldsymbol{8}$, which are schematically cataloged in the second and third sections of \hyperref[t5]{Table V}. The resulting list of explicit operators are displayed in Tables \hyperref[t8]{VIII} and \hyperref[t9]{IX}.

\renewcommand*{\arraystretch}{2}
\begin{table*}[t]\label{t8}
\begin{center}
 \begin{tabular}{|c||c|c|c|c|c|}
 \hline
\rule{0pt}{3.4ex}\multirow{2}{*}[-1ex]{\ \ $\boldsymbol{3} \otimes \boldsymbol{3} \otimes \boldsymbol{\bar{6}}$\ \ } & \multicolumn{3}{c|}{Singlet (Lorentz + $\mathcal{G}_{\text{SM}}$)} & \multirow{2}{*}[-1ex]{\ \ $L$\ \ } & \multirow{2}{*}[-1ex]{\ \ $Y$\ \ } \\[0.5ex]
\cline{2-4}
& Generic & Specific & \ Coupling\ \ & & \\[0.83ex]
\hline
\hline
\multirow{7}{*}[-3.5ex]{\ \emph{Scalar} $\Phi_s$} \rule{0pt}{3.5ex} & \multirow{6}{*}[-3ex]{\ $(qq')\Phi^{\dagger}$\ \ } & \ $\bt{K}_s^{\ \,ij}\, \Phi^{\dagger s}\,(\,\overbar{q^{\text{c}}_{\text{R}}}_{Ii}\,  q_{\text{R}Jj})$\ \ & $\lambda_{IJ}$ & \multirow{7}{*}[-3.5ex]{0} & \multirow{2}{*}[-1ex]{\ $\{-\tfrac{2}{3},\tfrac{1}{3},\tfrac{4}{3}\}$\ }\\[0.83ex]
\cline{3-4}
 \rule{0pt}{3.5ex} &  & \ \ $\bt{K}_s^{\ \,ij}\, \Phi^{\dagger s}\,(\,\overbar{q^{\text{c}}_{\text{R}}}_{Ii}\, \sigma^{\mu\nu} q_{\text{R}Jj})\,B_{\mu\nu}$\ \ & $\dfrac{1}{\Lambda_{\Phi}^2}\, \lambda_{IJ}$ &  & \\[0.83ex]
 \cline{3-4}
 \cline{6-6}
 \rule{0pt}{3.5ex} &  & \ $\bt{K}_s^{\ \,ij}\, \Phi^{\dagger s}\,(\,\overbar{Q^{\text{c}}_{\text{L}}}_{Ii}\,  \ii \tau^2\, Q_{\text{L}Jj})$\ \ & $\lambda_{IJ}$ &  & \multirow{4}{*}[-2ex]{$\tfrac{1}{3}$}\\[0.83ex]
 \cline{3-4} 
 \rule{0pt}{3.5ex} &  & \ $\bt{K}_s^{\ \,ij}\, \Phi^{\dagger s}\,(\,\overbar{Q^{\text{c}}_{\text{L}}}_{Ii}\, \sigma^{\mu\nu} \ii \tau^2\, Q_{\text{L}Jj})\,B_{\mu\nu}$\ \ & $\dfrac{1}{\Lambda_{\Phi}^2}\,\lambda_{IJ}$ &  & \\[0.83ex]
 \cline{3-4}
  \rule{0pt}{3.5ex} &  & \ $\bt{K}_s^{\ \,ij}\, \Phi^{\dagger s}\,(\,\overbar{Q^{\text{c}}_{\text{L}}}_{Ii}H H^{\dagger}Q_{\text{L}Jj})$\ \ & \multirow{3}{*}[-1.75ex]{$\dfrac{1}{\Lambda_{\Phi}^2}\,\lambda_{IJ}$} &  & \\[0.83ex]
 \cline{3-3} 
   \rule{0pt}{3.5ex} &  & \ $\bt{K}_s^{\ \,ij}\, \Phi^{\dagger s}\,(\,\overbar{Q^{\text{c}}_{\text{L}}}_{Ii}\,\sigma^{\mu\nu}\bt{t}_{\boldsymbol{2}}^A\, Q_{\text{L}Jj})\, W_{\mu\nu\,A}$\ \ &  &  & \\[0.83ex]
  \cline{2-3}
  \cline{6-6}
  \rule{0pt}{3.5ex} & \ $(qq')|H|^2\Phi^{\dagger}$\ \ & \ $\bt{K}_s^{\ \,ij}\, \Phi^{\dagger s}\,(\,\overbar{q^{\text{c}}_{\text{R}}}_{Ii}\,  q_{\text{R}Jj})\, |H|^2$\ \ &  & & $\{-\tfrac{2}{3},\tfrac{1}{3},\tfrac{4}{3}\}$\\[0.83ex]
 \hline
 \hline
\multirow{10}{*}[-5ex]{\ \emph{Dirac} $\Psi_s$} \rule{0pt}{3.5ex}  & \multirow{6}{*}[-3ex]{$(qq')(\bar{\Psi}\ell)$} & \ \ $\bt{K}_s^{\ \,ij}\, (\,\overbar{q^{\text{c}}_{\text{R}}}_{Ii}\, q_{\text{R}Jj})(\bar{\Psi}^s\,\ell_{\text{R}X})$\ \ & $\dfrac{1}{\Lambda_{\Psi}^2}\, \kappa_{IJ}^X$ & \multirow{8}{*}[-4ex]{1} & \multirow{3}{*}[-1.5ex]{$\ \{-\tfrac{5}{3},-\tfrac{2}{3},\tfrac{1}{3}\}$\ } \\[0.83ex]
\cline{3-4}
 \rule{0pt}{3.5ex} &  & \ \ $\bt{K}_s^{\ \,ij}\, (\,\overbar{q^{\text{c}}_{\text{R}}}_{Ii}\, q_{\text{R}Jj})(\bar{\Psi}^s H^{\dagger} L_{\text{L}X})$\ \ & \multirow{2}{*}[-1ex]{$\dfrac{1}{\Lambda_{\Psi}^3}\,\kappa_{IJ}^X$} &  & \\[0.83ex]
 \cline{3-3} 
 \rule{0pt}{3.5ex} &  & \ \ $\bt{K}_s^{\ \,ij}\, (\,\overbar{q^{\text{c}}_{\text{R}}}_{Ii}\, \sigma^{\mu\nu}q_{\text{R}Jj})(\bar{\Psi}^s\sigma_{\mu\nu} H^{\dagger} L_{\text{L}X})$\ \ &  &  & \\[0.83ex]
\cline{3-4}
\cline{6-6}
  \rule{0pt}{3.5ex} &  & \ \ $\bt{K}_s^{\ \,ij}\, (\,\overbar{Q^{\text{c}}_{\text{L}}}_{Ii}\,  \ii \tau^2\, Q_{\text{L}Jj})(\bar{\Psi}^s\, \ell_{\text{R}X})$\ \ & $\dfrac{1}{\Lambda_{\Psi}^2}\,\kappa_{IJ}^X$ &  & $\tfrac{4}{3}$\\[0.83ex]
 \cline{3-4} 
 \cline{6-6}
   \rule{0pt}{3.5ex} &  & \ \ $\bt{K}_s^{\ \,ij}\, (\,\overbar{Q^{\text{c}}_{\text{L}}}_{Ii}\,  \ii \tau^2\, Q_{\text{L}Jj})(\bar{\Psi}^s H^{\dagger}L_{\text{L}X})$\ \ & \multirow{2}{*}[-1ex]{$\dfrac{1}{\Lambda_{\Psi}^3}\,\kappa_{IJ}^X$} &  & $-\tfrac{2}{3}$\\[0.83ex]
 \cline{3-3} 
 \cline{6-6}
    \rule{0pt}{3.5ex} &  & \ \ $\bt{K}_s^{\ \,ij}\, (\,\overbar{Q^{\text{c}}_{\text{L}}}_{Ii} H \gamma^{\mu} q_{\text{R}Jj})(\bar{\Psi}^s\gamma_{\mu}\, \ell_{\text{R}X})$\ \ &  &  & $\{-\tfrac{2}{3},\tfrac{1}{3}\}$\\[0.83ex]
 \cline{2-4}
 \cline{6-6}
\rule{0pt}{3.5ex} & \multirow{2}{*}[-1ex]{$(\bar{\Psi}q)(q\ell)$} & $\bt{K}_s^{\ \,ij}\, (\bar{\Psi}^s \sigma^{\mu\nu}q_{\text{R}Ii})(\,\overbar{q^{\text{c}}_{\text{R}}}_{Jj}\,\sigma_{\mu\nu}\,\ell_{\text{R}X})$ & $\dfrac{1}{\Lambda_{\Psi}^3}\,\kappa_{IJ}^X$ &  & $\{-\tfrac{5}{3},-\tfrac{2}{3},\tfrac{1}{3}\}$ \\[0.83ex]
\cline{3-4}
\cline{6-6}
\rule{0pt}{3.5ex} & & $\bt{K}_s^{\ \,ij}\, (\bar{\Psi}^s \gamma^{\mu}q_{\text{R}Ii})(\,\overbar{Q^{\text{c}}_{\text{L}}}_{Jj}H\gamma_{\mu}\,\ell_{\text{R}X})$ & \multirow{3}{*}[-1.75ex]{$\dfrac{1}{\Lambda_{\Psi}^2}\,\kappa_{IJ}^X$} &  & $\{-\tfrac{2}{3},\tfrac{1}{3}\}$ \\[0.83ex]
\cline{2-3}
\cline{5-6}
\rule{0pt}{3.5ex} & \multirow{2}{*}[-1ex]{$(\bar{\Psi}q)(\bar{\ell}q)$} & $\bt{K}_s^{\ \,ij}\, (\bar{\Psi}^s \gamma^{\mu}q_{\text{R}Ii})(\bar{L}_{\text{L}X}\gamma_{\mu}\,\ii \tau^2\,Q_{\text{L}Jj})$ &  & \multirow{2}{*}[-1ex]{$-1$} & $\{\tfrac{1}{3},\tfrac{4}{3}\}$ \\[0.83ex]
\cline{3-3}
\cline{6-6}
\rule{0pt}{3.5ex} &  & $\bt{K}_s^{\ \,ij}\, (\bar{\Psi}^s \gamma^{\mu}q_{\text{R}Ii})(\bar{\ell}_{\text{R}X}\gamma_{\mu} q_{\text{R}Jj})$ &  &  & $\{\tfrac{1}{3},\tfrac{4}{3},\tfrac{7}{3}\}$ \\[0.83ex]
 \hline
 \end{tabular}\\[0.83ex]
 \end{center}
 \caption{Gauge-invariant operators coupling color sextets to quark pairs, based on the $\mathrm{SU}(3)$ invariant $\boldsymbol{3} \otimes \boldsymbol{3} \otimes \boldsymbol{\bar{6}}$ with Clebsch-Gordan coefficients $\bt{K}_s^{\ \,ij}$ (\emph{viz}. \hyperref[aA]{Appendix A}). Hermitian conjugates also exist where distinct. Quark generation indices $\{I,J\}$, lepton generation indices $\{X,Y\}$, and all color indices are kept explicit, while Dirac spinor indices and $\mathrm{SU}(2)_{\text{L}}$ indices are suppressed. $L$ and $Y$ are respectively the sextet lepton numbers and weak hypercharges in each scenario.}
 \end{table*}
\renewcommand*{\arraystretch}{1}

\renewcommand*{\arraystretch}{2}
\begin{table*}[t]\label{t9}
\begin{center}
 \begin{tabular}{|c||c|c|c|c|c|}
 \hline
\rule{0pt}{3.4ex}\multirow{2}{*}[-1ex]{\ \ $\boldsymbol{3} \otimes \boldsymbol{6} \otimes \boldsymbol{8}$\ \ } & \multicolumn{3}{c|}{Singlet (Lorentz + $\mathcal{G}_{\text{SM}}$)} & \multirow{2}{*}[-1ex]{\ \ $L$\ \ } & \multirow{2}{*}[-1ex]{\ \ $Y$\ \ } \\[0.5ex]
\cline{2-4}
& Generic & Specific & \ Coupling\ \ & & \\[0.83ex]
\hline
\hline
\multirow{2}{*}[-1ex]{\ \emph{Scalar} $\Phi_s$} \rule{0pt}{3.5ex} & \ $(q\ell)\Phi G$\ \ & \ $\bt{J}^{\,s\,ia}\, \Phi_s\,(\,\overbar{q^{\text{c}}_{\text{R}}}_{Ii}\, \sigma^{\mu\nu} \ell_{\text{R}X})\, G_{\mu\nu\,a}$\ \ & $\dfrac{1}{\Lambda_{\Phi}^2}\,\lambda^X_I$ & $-1$ & \,$\{\tfrac{1}{3},\tfrac{4}{3}\}$\ \\[0.83ex]
\cline{2-6}
 \rule{0pt}{3.5ex} & $(\bar{\ell}q)\Phi G$ & \ $\bt{J}^{\,s\,ia}\, \Phi_s\,(\bar{L}_{\text{L}X} H \sigma^{\mu\nu} q_{\text{R}Ii})\,G_{\mu\nu\,a}$\ \ & $\dfrac{1}{\Lambda_{\Phi}^3}\, \lambda_I^X$ & $1$ & \ $\{-\tfrac{5}{3},-\tfrac{2}{3}\}$\ \ \\[0.83ex]
 \hline
 \hline
\multirow{3}{*}[-1.75ex]{\ \emph{Dirac} $\Psi_s$} \rule{0pt}{3.5ex}  & \multirow{2}{*}[-1ex]{$(q\Psi)G$} & $\bt{J}^{\,s\,ia}\, (\,\overbar{q^{\text{c}}_{\text{R}}}_{Ii}\,\sigma^{\mu\nu}\Psi_s)\,G_{\mu\nu\,a}$ & $\dfrac{1}{\Lambda_{\Psi}}\, \kappa_I$ & \multirow{3}{*}[-1.75ex]{0} & \multirow{3}{*}[-1.75ex]{$\{-\tfrac{2}{3},\tfrac{1}{3}\}$} \\[0.83ex]
\cline{3-4}
\rule{0pt}{3.5ex} & & \ $\bt{J}^{\,s\,ia}\, (\,\overbar{q^{\text{c}}_{\text{R}}}_{Ii}\,\Psi_s)\,B^{\mu\nu}G_{\mu\nu\,a}$\ \ & \multirow{2}{*}[-1ex]{$\dfrac{1}{\Lambda_{\Psi}^3}\,\kappa_I$} & & \\[0.83ex]
\cline{2-3}
\rule{0pt}{3.5ex} & \ $(q\Psi)|H|^2G$\ \ & $\bt{J}^{\,s\,ia}\, (\,\overbar{q^{\text{c}}_{\text{R}}}_{Ii}\,\sigma^{\mu\nu}\Psi_s)\,|H|^2\, G_{\mu\nu\,a}$ &  & & \\[0.83ex]
 \hline
 \end{tabular}\\[0.83ex]
 \end{center}
 \caption{Gauge-invariant operators coupling color sextets to quarks and gluons, based on the $\mathrm{SU}(3)$ invariant $\boldsymbol{3} \otimes \boldsymbol{6} \otimes \boldsymbol{8}$ with Clebsch-Gordan coefficients $\bt{J}^{\,s\,ia}$ (\emph{viz}. \hyperref[aA]{Appendix A}). Conventions are similar to those of \hyperref[t8]{Table VIII}.}
 \end{table*}
\renewcommand*{\arraystretch}{1}

These tables show all the details hidden in the schematic operator lists by fully specifying the wide range of operators with color indices, Clebsch-Gordan coefficients, couplings\footnote{In the interest of generality, the scalar couplings $\lambda_{I(J)}^{(X)}$ and Dirac couplings $\kappa_{I(J)}^{(X)}$ are matrices in quark (and sometimes lepton) generation space.}, and EFT cutoffs $\Lambda_{\{\Phi,\Psi\}}$ made explicit. In addition, these tables specify the lepton numbers $L$ and weak hypercharges $Y$ the sextet field must assume in order to preserve gauge invariance and the accidental lepton number conservation of the Standard Model. $\mathrm{SU}(2)_{\text{L}}$ (weak) indices are suppressed throughout, and Dirac indices are contracted between objects within parentheses.

\hyperref[t8]{Table VIII}, which targets the invariant $\boldsymbol{3} \otimes \boldsymbol{3} \otimes \boldsymbol{\bar{6}}$, is fairly large, even though there exists only one way to contract color indices to form this invariant.  This structure minimally couples a color sextet to quark pairs $qq'$, which historically motivated the term ``sextet diquarks''. The first and third rows of \hyperref[t8]{Table VIII} reproduce the gauge-invariant interactions cataloged for weak-singlet color-sextet scalars in \cite{Shu:2009xf,Han_2010_2}. In addition to these known couplings, we find numerous operators with various quark chiralities and extra SM gauge and Higgs bosons. The operators become even more exotic for sextet fermions in this color structure, with leptons being necessary in every case to form gauge and Lorentz singlets.

\hyperref[t9]{Table IX} concerns the invariant $\boldsymbol{3} \otimes \boldsymbol{6} \otimes \boldsymbol{8}$, which despite being a simple three-field invariant has received scant attention in the literature. This invariant couples a color sextet to a quark and a gluon, and while no operators can be built at mass dimension four, it is potentially very important in LHC searches for color sextets. Here, as for the other three-field invariant, we find basic $qg$ couplings, as well as interactions containing extra $B$ or Higgs bosons. Interestingly, the situation with respect to leptons is flipped relative to \hyperref[t8]{Table VIII}, with the scalar sextet interactions requiring leptons. In short, this table depicts a fairly minimal but rich portal between the Standard Model and color-sextet scalars and fermions. We highlight some of the operators in this table in a phenomenological investigation in \hyperref[s3]{Section III}.
\section{Sextets at the LHC}
\label{s3}

In the previous section we introduced a wide variety of operators governing the production and decay of exotic color-charged states in the six-dimensional representation ($\boldsymbol{6}$) of the Standard Model $\mathrm{SU}(3)_{\text{c}}$. In this section we exploit the operator catalog to investigate models of color-sextet fields based on a subset of these operators. In particular, we consider color-sextet fermions and scalars coupling to gluons, up- or down-type quarks, and sometimes leptons and the $\mathrm{U}(1)_Y$ gauge boson(s) $B$; these couplings are enabled by the color invariant $\boldsymbol{3} \otimes \boldsymbol{6} \otimes \boldsymbol{8}$. The sextet fermion models are defined by
\begin{multline}\label{sFmodel}
    \mathcal{L} \supset \bar{\Psi}_q(\ii \slashed{D} - m_{\Psi_q})\Psi_q\\ + \frac{1}{\Lambda_{\Psi_q}}\,[\kappa_q^I\,\bt{J}^{\,s\,ia}\,(\,\overbar{q^{\text{c}}_{\text{R}}}_{Ii}\,\sigma^{\mu\nu}\Psi_{qs})\, G_{\mu\nu\,a} + \text{H.c.}]\\ + \frac{1}{\Lambda_{\Psi_{qB}}^3}\,[\kappa_{qB}^I\,\bt{J}^{\,s\,ia}\,(\,\overbar{q^{\text{c}}_{\text{R}}}_{Ii}\,\Psi_{qs})\, B^{\mu\nu}G_{\mu\nu\,a} + \text{H.c.}]
\end{multline}
for $q \in \{u,d\}$ (so for instance $\Psi_u$ couples to up-type quarks). The models for sextet scalars are analogously given by
\begin{multline}\label{sSmodel}
    \mathcal{L} \supset (D_{\mu}\Phi_q)^{\dagger}D^{\mu}\Phi_q - m_{\Phi_q}^2 \Phi_q^{\dagger}\Phi_q\\ + \frac{1}{\Lambda^2_{\Phi_q}}\,[\lambda^{XI}_q \bt{J}^{\,s\, ia}\,\Phi_{qs}\,(\,\overbar{q^{\text{c}}_{\text{R}}}_{Ii}\,\sigma^{\mu\nu}\ell_{\text{R}X})\, G_{\mu\nu\,a} + \text{H.c.}].
\end{multline}
In both Lagrange densities \eqref{sFmodel} and \eqref{sSmodel}, as elsewhere, spinor indices are contracted within parentheses. The couplings and cutoff scales are taken with light modification from \hyperref[t9]{Table IX}. The Clebsch-Gordan coefficients $\bt{J}^{\,s\,ia}$ (with Hermitian conjugates $\bar{\bt{J}}_{s\, ai}$) providing the gauge-invariant contraction of a color sextet with a quark and a gluon were introduced in \hyperref[t1]{Table I}. We discuss these novel coefficients in detail and explicitly provide them in a useful basis in \hyperref[aA]{Appendix A}. As we noted in \hyperref[s2]{Section II}, these color sextets must have particular weak hypercharges and lepton numbers in order to preserve the symmetries of the Standard Model. We summarize the quantum numbers of the fields in \eqref{sFmodel} and \eqref{sSmodel} in \hyperref[fields]{Table X}.

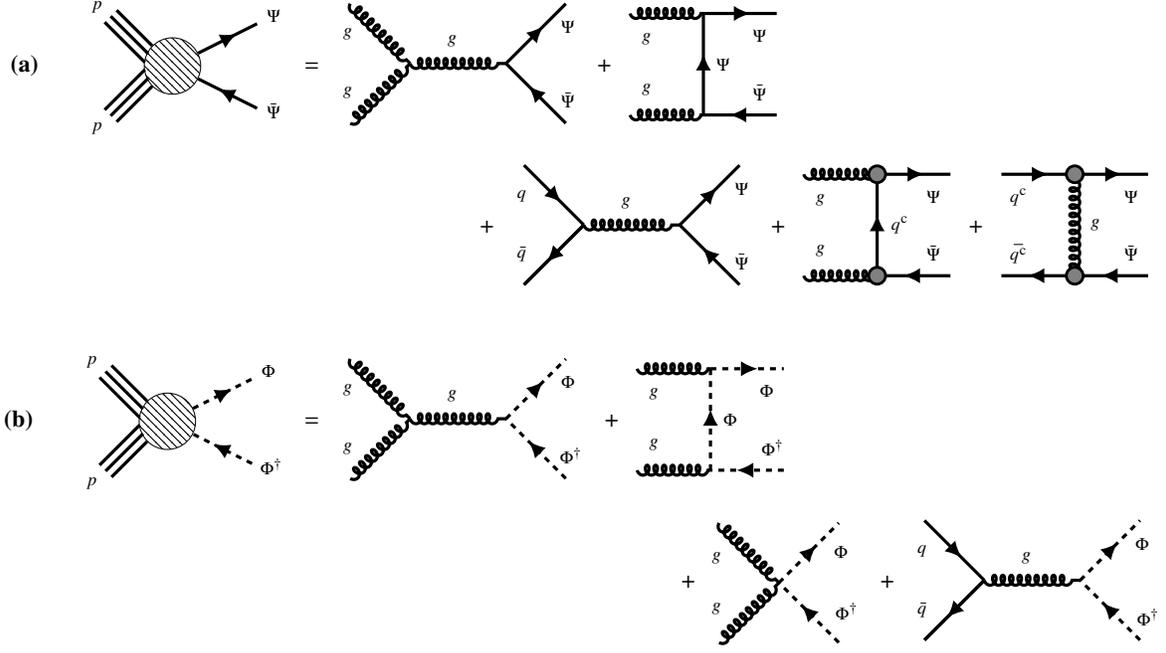
\begin{figure*}[t]\label{ppdiagrams}
\begin{align*}
\textbf{(a)}\ \ \ \ \ \ \ \ \scalebox{0.75}{\begin{tikzpicture}[baseline={([yshift=-.5ex]current bounding box.center)},xshift=12cm]
\begin{feynman}[large]
\vertex (i1);
\vertex [right = 1.25cm of i1, blob] (i2){};
\vertex [below left = 0.15cm of i2] (q1);
\vertex [above left=0.46 cm of q1] (r1);
\vertex [below left =0.15cm of q1] (q2);
\vertex [above left = 0.39cm of q2] (r2);
\vertex [above left=0.15cm of i2] (l1);
\vertex [below left = 0.46cm of l1] (r3);
\vertex [above left=0.15cm of l1] (l2);
\vertex [below left=0.39cm of l2] (r4);
\vertex [above=0.2cm of i2] (q3);
\vertex [above=0.2cm of q3] (q4);
\vertex [above left=1.4cm of i2] (v3);
\vertex [below left=0.15cm of v3] (p1);
\vertex [below left =0.15cm of p1] (p2);
\vertex [below left=1.4cm of i2] (v4);
\vertex [above left=0.15cm of v4] (p3);
\vertex [above left =0.15cm of p3] (p4);
\vertex [above right=0.75 cm and 1.5cm of i2] (v1);
\vertex [below right=0.75cm and 1.5cm of i2] (v2);
\diagram* {
(p1) -- [ultra thick] (r1),
(p2) -- [ultra thick] (r2),
(p3) -- [ultra thick] (r3),
(p4) -- [ultra thick] (r4),
(v3) -- [ultra thick] (i2),
(v4) -- [ultra thick] (i2),
(i2) -- [ultra thick, fermion ] (v1),
(v2) -- [ultra thick, fermion] (i2),
};
\end{feynman}
\node at (0.425,1.05) {$p$};
\node at (0.425,-1.1) {$p$};
\node at (3.55,0.9) {$\Psi$};
\node at (3.55,-0.8) {$\bar{\Psi}$};
\end{tikzpicture}}\  \ &=\  \ \begin{multlined}[t][10.8cm]\scalebox{0.75}{\begin{tikzpicture}[baseline={([yshift=-0.9ex]current bounding box.center)},xshift=12cm]
\begin{feynman}[large]
\vertex (i1);
\vertex [right = 1.7cm of i1] (i2);
\vertex [above left=1.5 cm of i1] (v1);
\vertex [below left=1.5cm of i1] (v2);
\vertex [above right=1.5cm of i2] (v3);
\vertex [below right=1.5cm of i2] (v4);
\diagram* {
(i1) -- [ultra thick, gluon] (i2),
(v1) -- [ultra thick, gluon] (i1),
(v2) -- [ultra thick, gluon] (i1),
(i2) -- [ultra thick, fermion] (v3),
(v4) -- [ultra thick, fermion] (i2),
};
\end{feynman}
\node at (-1.1,0.52) {$g$};
\node at (-1.1,-0.52) {$g$};
\node at (0.75,0.4) {$g$};
\node at (2.8,0.65) {$\Psi$};
\node at (2.8,-0.65) {$\bar{\Psi}$};
\end{tikzpicture}}\ \ +\ \ \scalebox{0.75}{\begin{tikzpicture}[baseline={([yshift=-0.9ex]current bounding box.center)},xshift=12cm]
\begin{feynman}[large]
\vertex (i1);
\vertex [below = 1.8cm of i1] (i2);
\vertex [left = 1.3cm of i1] (v1);
\vertex [right= 1.3cm of i1] (v2);
\vertex [left = 1.3cm of i2] (v3);
\vertex [right=1.3cm of i2] (v4);
\diagram*{
(i2) -- [ultra thick, fermion] (i1),
(v1) -- [ultra thick, gluon] (i1),
(i1) -- [ultra thick, fermion] (v2),
(v3) -- [ultra thick, gluon] (i2),
(v4) -- [ultra thick, fermion] (i2),
};
\end{feynman}
\node at (-1,-0.45) {$g$};
\node at (-1,-1.35) {$g$};
\node at (0.35,-0.9) {$\Psi$};
\node at (1.,-0.4) {$\Psi$};
\node at (1,-1.4) {$\bar{\Psi}$};
\end{tikzpicture}}\\[2.5ex] +\ \ \scalebox{0.75}{\begin{tikzpicture}[baseline={([yshift=-0.9ex]current bounding box.center)},xshift=12cm]
\begin{feynman}[large]
\vertex (i1);
\vertex [right = 1.7cm of i1] (i2);
\vertex [above left=1.5 cm of i1] (v1);
\vertex [below left=1.5cm of i1] (v2);
\vertex [above right=1.5cm of i2] (v3);
\vertex [below right=1.5cm of i2] (v4);
\diagram* {
(i1) -- [ultra thick, gluon] (i2),
(v1) -- [ultra thick, fermion] (i1),
(i1) -- [ultra thick, fermion] (v2),
(v2) -- [ultra thick] (i1),
(i2) -- [ultra thick, fermion] (v3),
(v4) -- [ultra thick, fermion] (i2),
};
\end{feynman}
\node at (-1.1,0.52) {$q$};
\node at (-1.1,-0.52) {$\bar{q}$};
\node at (0.75,0.4) {$g$};
\node at (2.8,0.65) {$\Psi$};
\node at (2.8,-0.65) {$\bar{\Psi}$};
\end{tikzpicture}}\ \ +\ \ \scalebox{0.75}{\begin{tikzpicture}[baseline={([yshift=-0.9ex]current bounding box.center)},xshift=12cm]
\begin{feynman}[large]
\vertex (i1);
\vertex [below = 1.8cm of i1] (i2);
\vertex [left = 1.3cm of i1] (v1);
\vertex [right= 1.3cm of i1] (v2);
\vertex [left = 1.3cm of i2] (v3);
\vertex [right=1.3cm of i2] (v4);
\diagram*{
(i2) -- [ultra thick, fermion] (i1),
(v1) -- [ultra thick, gluon] (i1),
(i1) -- [ultra thick, fermion] (v2),
(v3) -- [ultra thick, gluon] (i2),
(v4) -- [ultra thick, fermion] (i2),
};
\end{feynman}
\node at (-1,-0.45) {$g$};
\node at (-1,-1.35) {$g$};
\node at (0.4,-0.9) {$q^{\text{c}}$};
\node at (1.,-0.4) {$\Psi$};
\node at (1,-1.4) {$\bar{\Psi}$};
\node at (0,0) [circle,draw=black,line width=0.5mm, fill=gray,inner sep=3pt]{};
\node at (0,-1.8) [circle,draw=black,line width=0.5mm, fill=gray,inner sep=3pt]{};
\end{tikzpicture}}\ \ +\ \ \scalebox{0.75}{\begin{tikzpicture}[baseline={([yshift=-0.9ex]current bounding box.center)},xshift=12cm]
\begin{feynman}[large]
\vertex (i1);
\vertex [below = 1.8cm of i1] (i2);
\vertex [left = 1.3cm of i1] (v1);
\vertex [right= 1.3cm of i1] (v2);
\vertex [left = 1.3cm of i2] (v3);
\vertex [right=1.3cm of i2] (v4);
\diagram*{
(i2) -- [ultra thick, gluon] (i1),
(v1) -- [ultra thick, fermion] (i1),
(i1) -- [ultra thick, fermion] (v2),
(i2) -- [ultra thick, fermion] (v3),
(v4) -- [ultra thick, fermion] (i2),
};
\end{feynman}
\node at (-1,-0.4) {$q^{\text{c}}$};
\node at (-1,-1.4) {$\overbar{q^{\text{c}}}$};
\node at (0.37,-0.9) {$g$};
\node at (1.,-0.4) {$\Psi$};
\node at (1,-1.4) {$\bar{\Psi}$};
\node at (0,0) [circle,draw=black,line width=0.5mm, fill=gray,inner sep=3pt]{};
\node at (0,-1.8) [circle,draw=black,line width=0.5mm, fill=gray,inner sep=3pt]{};
\end{tikzpicture}}\end{multlined}\\[5ex]
\textbf{(b)}\ \ \ \ \ \ \ \ \scalebox{0.75}{\begin{tikzpicture}[baseline={([yshift=-.5ex]current bounding box.center)},xshift=12cm]
\begin{feynman}[large]
\vertex (i1);
\vertex [right = 1.25cm of i1, blob] (i2){};
\vertex [below left = 0.15cm of i2] (q1);
\vertex [above left=0.46 cm of q1] (r1);
\vertex [below left =0.15cm of q1] (q2);
\vertex [above left = 0.39cm of q2] (r2);
\vertex [above left=0.15cm of i2] (l1);
\vertex [below left = 0.46cm of l1] (r3);
\vertex [above left=0.15cm of l1] (l2);
\vertex [below left=0.39cm of l2] (r4);
\vertex [above=0.2cm of i2] (q3);
\vertex [above=0.2cm of q3] (q4);
\vertex [above left=1.4cm of i2] (v3);
\vertex [below left=0.15cm of v3] (p1);
\vertex [below left =0.15cm of p1] (p2);
\vertex [below left=1.4cm of i2] (v4);
\vertex [above left=0.15cm of v4] (p3);
\vertex [above left =0.15cm of p3] (p4);
\vertex [above right=0.75 cm and 1.5cm of i2] (v1);
\vertex [below right=0.75cm and 1.5cm of i2] (v2);
\diagram* {
(p1) -- [ultra thick] (r1),
(p2) -- [ultra thick] (r2),
(p3) -- [ultra thick] (r3),
(p4) -- [ultra thick] (r4),
(v3) -- [ultra thick] (i2),
(v4) -- [ultra thick] (i2),
(i2) -- [ultra thick, charged scalar ] (v1),
(v2) -- [ultra thick, charged scalar] (i2),
};
\end{feynman}
\node at (0.425,1.05) {$p$};
\node at (0.425,-1.1) {$p$};
\node at (3.55,0.9) {$\Phi$};
\node at (3.6,-0.8) {$\Phi^{\dagger}$};
\end{tikzpicture}}\  \,&=\  \ \begin{multlined}[t][11cm]\scalebox{0.75}{\begin{tikzpicture}[baseline={([yshift=-0.9ex]current bounding box.center)},xshift=12cm]
\begin{feynman}[large]
\vertex (i1);
\vertex [right = 1.7cm of i1] (i2);
\vertex [above left=1.5 cm of i1] (v1);
\vertex [below left=1.5cm of i1] (v2);
\vertex [above right=1.5cm of i2] (v3);
\vertex [below right=1.5cm of i2] (v4);
\diagram* {
(i1) -- [ultra thick, gluon] (i2),
(v1) -- [ultra thick, gluon] (i1),
(v2) -- [ultra thick, gluon] (i1),
(i2) -- [ultra thick, charged scalar] (v3),
(v4) -- [ultra thick, charged scalar] (i2),
};
\end{feynman}
\node at (-1.1,0.52) {$g$};
\node at (-1.1,-0.52) {$g$};
\node at (0.75,0.4) {$g$};
\node at (2.8,0.65) {$\Phi$};
\node at (2.85,-0.65) {$\Phi^{\dagger}$};
\end{tikzpicture}}\ \ +\ \ \scalebox{0.75}{\begin{tikzpicture}[baseline={([yshift=-0.9ex]current bounding box.center)},xshift=12cm]
\begin{feynman}[large]
\vertex (i1);
\vertex [below = 1.8cm of i1] (i2);
\vertex [left = 1.3cm of i1] (v1);
\vertex [right= 1.3cm of i1] (v2);
\vertex [left = 1.3cm of i2] (v3);
\vertex [right=1.3cm of i2] (v4);
\diagram*{
(i2) -- [ultra thick, charged scalar] (i1),
(v1) -- [ultra thick, gluon] (i1),
(i1) -- [ultra thick, charged scalar] (v2),
(v3) -- [ultra thick, gluon] (i2),
(v4) -- [ultra thick, charged scalar] (i2),
};
\end{feynman}
\node at (-1,-0.45) {$g$};
\node at (-1,-1.35) {$g$};
\node at (0.35,-0.9) {$\Phi$};
\node at (1.,-0.4) {$\Phi$};
\node at (1.1,-1.4) {$\Phi^{\dagger}$};
\end{tikzpicture}}\\[2.5ex] +\ \scalebox{0.75}{\begin{tikzpicture}[baseline={([yshift=-.5ex]current bounding box.center)},xshift=12cm]
\begin{feynman}[large]
\vertex (i1);
\vertex [above left = 1.5cm of i1] (g1);
\vertex [below left = 1.5cm of i1] (g2);
\vertex [above right=1.5 cm of i1] (v1);
\vertex [below right=1.5cm of i1] (v2);
\diagram* {
(g1) -- [ultra thick, gluon] (i1),
(g2) -- [ultra thick, gluon] (i1),
(i1) -- [ultra thick, charged scalar] (v1),
(v2) -- [ultra thick, charged scalar] (i1),
};
\end{feynman}
\node at (-1.1,0.45) {$g$};
\node at (-1.1,-0.45) {$g$};
\node at (1.1,0.55) {$\Phi$};
\node at (1.2,-0.5) {$\Phi^{\dagger}$};
\end{tikzpicture}}\ \ +\ \ \scalebox{0.75}{\begin{tikzpicture}[baseline={([yshift=-0.9ex]current bounding box.center)},xshift=12cm]
\begin{feynman}[large]
\vertex (i1);
\vertex [right = 1.7cm of i1] (i2);
\vertex [above left=1.5 cm of i1] (v1);
\vertex [below left=1.5cm of i1] (v2);
\vertex [above right=1.5cm of i2] (v3);
\vertex [below right=1.5cm of i2] (v4);
\diagram* {
(i1) -- [ultra thick, gluon] (i2),
(v1) -- [ultra thick, fermion] (i1),
(i1) -- [ultra thick, fermion] (v2),
(v2) -- [ultra thick] (i1),
(i2) -- [ultra thick, charged scalar] (v3),
(v4) -- [ultra thick, charged scalar] (i2),
};
\end{feynman}
\node at (-1.1,0.52) {$q$};
\node at (-1.1,-0.52) {$\bar{q}$};
\node at (0.75,0.4) {$g$};
\node at (2.8,0.65) {$\Phi$};
\node at (2.9,-0.65) {$\Phi^{\dagger}$};
\end{tikzpicture}}
\end{multlined}
\end{align*}
    \caption{Representative diagrams for pair production of color-sextet \textbf{(a)} fermions and \textbf{(b)} scalars. Blobs mark vertices corresponding to an effective operator with some cutoff scale. Quarks coupling directly to sextets must have appropriate hypercharge; \emph{viz}. \eqref{sFmodel} and \eqref{sSmodel}. These contributions are negligible for realistic $\Lambda_{\Psi_u},\Lambda_{\Psi_d}$.}
\end{figure*}

\renewcommand*{\arraystretch}{1.5}
\begin{table}\label{fields}
\begin{center}
 \begin{tabular}{|c||c|c|c|}
 \hline
\rule{0pt}{3.4ex} & \multicolumn{1}{c|}{} & \multicolumn{2}{c|}{\ Quantum numbers\ \ }\\[0.5ex]
\hline
\hline
\rule{0pt}{3.5ex} &  & $\mathcal{G}_{\text{SM}}$ & $L$\\[0.5ex]
\hline
   \multirow{2}{*}[-0.4ex]{\rotatebox[origin=c]{90}{\rule{0pt}{2.5ex} Scalars }} \rule{0pt}{3.5ex} & $\Phi_{u}$ & \ $(\boldsymbol{6},\boldsymbol{1},\tfrac{1}{3})$\ \ & \multirow{2}{*}[-1ex]{$-1$} \\[0.83ex]
\cline{2-3}
 \rule{0pt}{3.5ex} & $\Phi_{d}$ & $(\boldsymbol{6},\boldsymbol{1},\tfrac{4}{3})$ & \\[0.83ex]
 \hline
   \multirow{2}{*}[0.5ex]{\rotatebox[origin=c]{90}{\rule{0pt}{2.5ex}\ \ Fermions }} \rule{0pt}{3.5ex} & $\Psi_{u}$ & \ $(\boldsymbol{6},\boldsymbol{1},-\tfrac{2}{3})$\ \ & \multirow{2}{*}[-1ex]{$0$} \\[0.83ex]
\cline{2-3}
 \rule{0pt}{3.5ex} & $\Psi_{d}$ & $(\boldsymbol{6},\boldsymbol{1},\tfrac{1}{3})$ & \\[0.83ex]
 \hline
 \end{tabular}\\[0.83ex]
 \end{center}
 \caption{Exotic field content of color-sextet models considered in \hyperref[s3]{Section III}. Representations in SM gauge group $\mathcal{G}_{\text{SM}}$ and charges under accidental symmetries are noted.}
 \end{table}
\renewcommand*{\arraystretch}{1}

We implement these simplified models in version 2.3.43 of \textsc{FeynRules} \cite{FR_OG,FR_2}, a package for \textsc{Mathematica}$^\copyright$\ version 12.0 \cite{Mathematica}. Some notes on the implementation of the Clebsch-Gordan coefficients $\bt{J}^{\,s\,ia}$ are offered in \hyperref[aA]{Appendix A}. We have used \textsc{FeynRules} to generate a Universal FeynRules Output (UFO) \cite{UFO} for leading-order (LO) event generation in \textsc{MadGraph5\texttt{\textunderscore}aMC@NLO} (\textsc{MG5\texttt{\textunderscore}aMC}) version 3.2.0 \cite{MG5,MG5_EW_NLO}. For both cross section computations and event simulation, hard-scattering amplitudes have been convolved with the NNPDF\,2.3 LO set of parton distribution functions \cite{nnpdf}. The renormalization and factorization scales have been set to $\mu_{\text{R}} = \mu_{\text{F}} = m_{\Psi}\ \text{or}\ m_{\Phi}$.

\subsection{Cross sections and LHC signatures}\label{s3.1}

\begin{figure}\label{f2}
\includegraphics[scale=0.7]{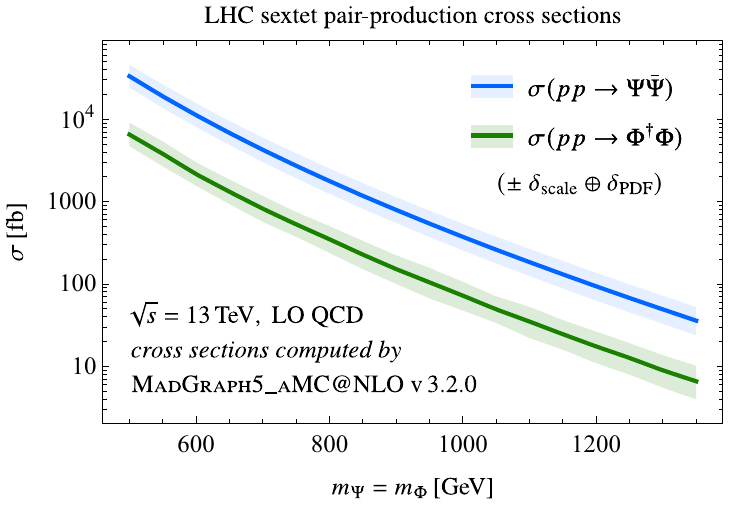}
\caption{Leading order cross sections for pair production of color-sextet fermions and scalars at the LHC as a function of sextet mass.}
\end{figure}

At a high-energy hadron collider, color sextets can be produced in pairs (predominantly through their $\mathrm{SU}(3)_{\text{c}}$ gauge interactions) as well as singly, often in association with SM leptons or gauge bosons. Representative diagrams for pair production are displayed in \hyperref[ppdiagrams]{Figure 1} and proceed \emph{via} gluon fusion and quark-antiquark annihilation.
Contributions to $\Psi\bar{\Psi}$ production from the higher dimensional operators in \eqref{sFmodel} are typically negligible for reasonable choices of $\Lambda_{\Psi_q}$. 
The cross sections of sextet fermion and scalar pair production at the LHC with $\sqrt{s}=13\,\text{TeV}$ are displayed in \hyperref[f2]{Figure 2},
with systematic errors estimated by adding the scale and PDF variations reported by \textsc{MG5\texttt{\textunderscore}aMC} in quadrature.
Because pair production is dominated by gauge interactions, the cross sections are not only essentially independent of $\kappa_q^I$ and $\Lambda_{\Psi_q}$, but are also nearly identical for sextets coupling to up- and down-type quarks. The spin of the sextet is important, with the scalar cross section(s) hovering a bit less than an order of magnitude below those of the fermions. All cross sections are of $\mathcal{O}(1\text{--}10)\,\text{pb}$ for masses below the TeV scale but fall quite steeply with increasing sextet mass. It is worth noting that these results are consistent with the small existing literature for sextet fermions \cite{Chivukula_91,Celikel_98} and scalars \cite{Han_2010,Han_2010_2,PhysRevD.79.054002}.

\begin{figure}\label{f3}
\includegraphics[scale=0.685]{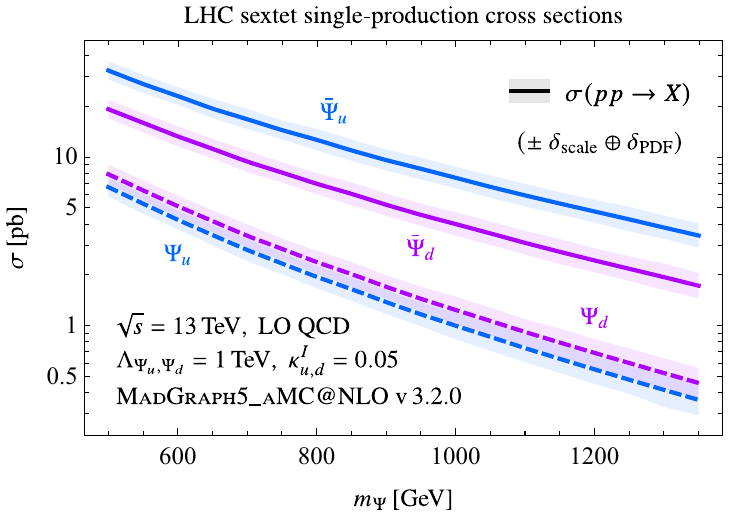}
\caption{Cross sections of color-sextet fermion single production. These are comparable to fermion pair-production cross sections for indicated couplings/cutoffs. Conjugate fermion ($\bar{\Psi}$) cross sections dominate because quarks have greater parton luminosity than antiquarks at LHC.}
\end{figure}

The novel interactions between sextet fermions, quarks, and gluons in the second line of \eqref{sFmodel} allow sextet fermions to be singly produced in quark-gluon fusion. Here the quantum numbers of the sextet are significant, and we display the cross sections for both sextet fermions and their antiparticles in \hyperref[f3]{Figure 3}.
We have chosen a simple benchmark in which all first- and second-generation quarks couple to sextets with equal strength: $\kappa_u^I = \kappa_d^I = 0.05\ \forall\, I\in \{1,2\}$ and $\Lambda_{\Psi_u} = \Lambda_{\Psi_d} = 1\,\text{TeV}$. These choices correspond to single-production cross sections for all sextet fermions which are comparable to those for pair production. The differences in cross section for the two fermions and their conjugates are largely the result of the significant difference between quark and antiquark parton luminosities at LHC. (One can see in \eqref{sFmodel} that the initial state for $\bar{\Psi}_q$ production is essentially $qg$, whereas $\Psi_q$ is produced by $\bar{q}g$.) The situation would be quite different at a $p\bar{p}$ collider like the Tevatron, but here the differences amount to factors of around five. As expected, these cross sections fall more gently with increasing $m_{\Psi}$ than that of pair production. 

\begin{figure}\label{f4}
\includegraphics[scale=0.7]{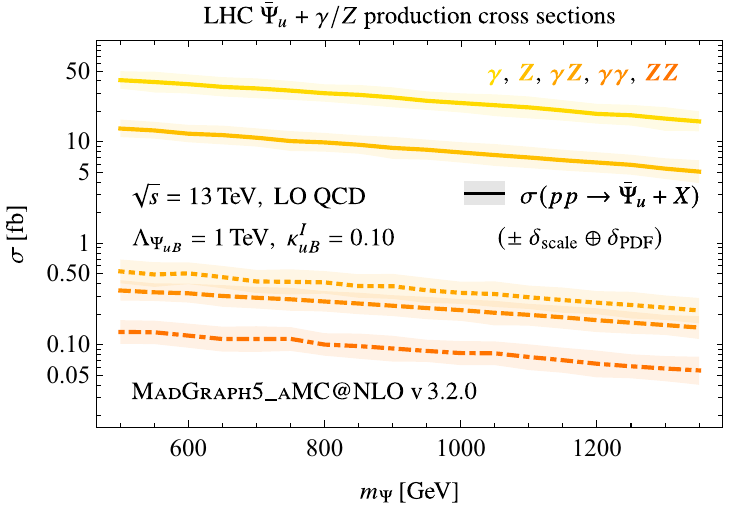}
\caption{Cross sections of $\bar{\Psi}_u$ single production in association with up to two photons and/or $Z$ bosons. Cross sections for $\Psi_u$ and $\Psi_d,\bar{\Psi}_d$ are smaller in analogy with single production (\emph{viz}. \hyperref[f2]{Figure 2} and discussion).}
\end{figure}

A third interesting production mode involving a sextet fermion is enabled by the final line of \eqref{sFmodel}. We find that processes with up to two photons or $Z$ bosons in the final state may have cross sections of $\mathcal{O}(1\text{--}10)\,\text{fb}$, which is on the margins of what is observable at the LHC. We display these cross sections for the optimal case of a sextet antifermion coupling to up-type quarks in \hyperref[f4]{Figure 4}.
These results correspond to a benchmark with (light) flavor-universal couplings ($\kappa_{uB}^I = 0.10\ \forall\, I\in\{1,2\}$) and a cutoff scale of $\Lambda_{uB} = 1\,\text{TeV}$. As expected, the cross section for $pp \to \bar{\Psi}_u + \gamma$ is a few times larger than that of $pp \to \bar{\Psi}_u + Z$. On the other hand, $\bar{\Psi}_u + \gamma Z$ is the largest of the two-boson associated production modes, followed by $\gamma\gamma$ and $ZZ$. For the same reasons as for unaccompanied single fermion production, other sextet fermions produced in association with photons and $Z$ bosons exhibit smaller cross sections.

\begin{figure}\label{f5}
\includegraphics[scale=0.7]{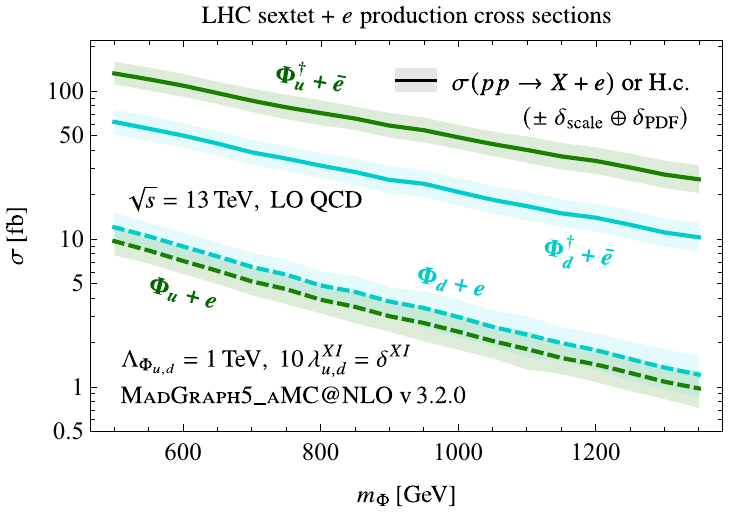}
\caption{Cross sections of sextet scalar single production in association with an electron or positron. Conjugate scalar production dominates in a fashion similar to sextet fermion single production. Cross sections for associated production with $\mu,\tau$ are orders of magnitude smaller if all else is equal.}
\end{figure}

While no operator in \eqref{sSmodel} allows for the production of a lone color-sextet scalar at LHC, the production of such particles in association with SM leptons is allowed by the second line. These again are due to quark-gluon fusion and exhibit the same behavior with respect to $q/\bar{q}$ initial states. We display in \hyperref[f5]{Figure 5} some suggestive cross sections for sextet scalar production with an electron or its antiparticle. 
Here we adopt a benchmark in which the coupling matrices $\lambda^{XI}_{u,d}$ are diagonal in generation space, so for instance $\Psi_u + e$ is produced only by $\bar{u}g$ fusion. We specifically let $\lambda^{XI}_{u,d} = 0.1 \times \delta^{XI}$ and again choose cutoffs of $\Lambda_{\Phi_u} = \Lambda_{\Phi_d} = 1\,\text{TeV}$. In this scenario, these cross sections are up to a few times larger than those of sextet fermion with $\gamma/Z$ production.

\subsection{Constraining color sextets at LHC}\label{s3.2}

As we have seen, color-sextet particles can be produced at the LHC singly and in pairs, sometimes in association with leptons and bosons. They will subsequently decay to a variety of two or more SM particles. Many of these decay products will hadronize in a detector, ultimately producing final states with possibly large jet multiplicities, possibly accompanied by leptons and electrically neutral bosons. This rich phenomenology makes color-sextet models ripe for exploration at the LHC, and indeed a small subset of these models, mostly corresponding to the top row of \hyperref[t8]{Table VIII}, have received attention in the literature \cite{Han_2010,Han_2010_2,PhysRevD.77.011701,PhysRevD.87.115019}.

However, most of the important signatures generated by the example models \eqref{sFmodel} and \eqref{sSmodel} are fringe cases that have not been directly targeted by either experimental collaboration. This ability to evade LHC searches by producing exotic signatures is typical of our expanded color-sextet catalog; \emph{viz}. Tables \hyperref[t8]{VIII} and \hyperref[t9]{IX}. In this discussion, we enumerate the most interesting signatures worthy of future study, which in principle arise both from leading-order color-sextet decays to SM particles \cite{CMS:2018s1,ext_2018} and from sextet loop contributions to meson-antimeson mixing \cite{PhysRevD.87.115019}.

\begin{figure}\label{f6}
\includegraphics[scale=0.67]{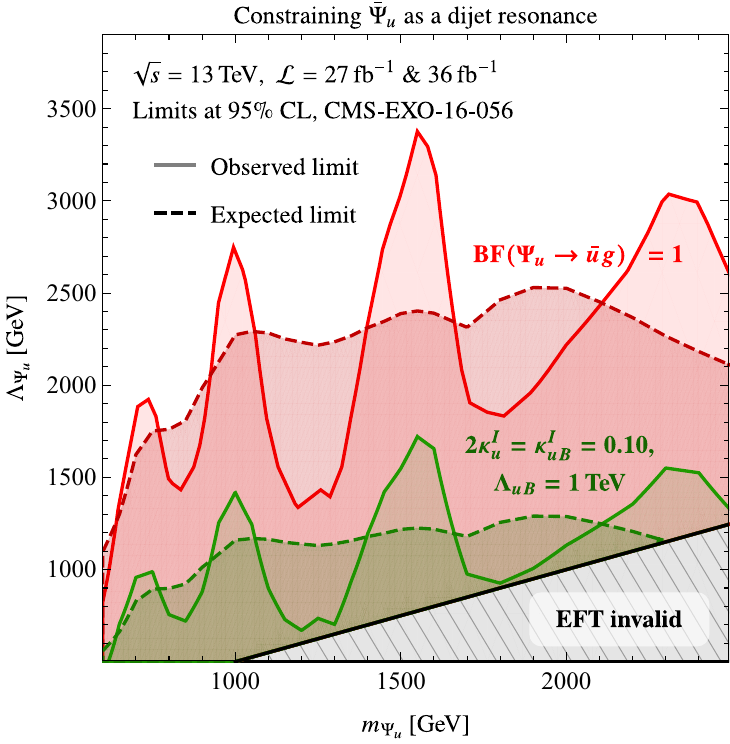}
\caption{Parameter space excluded for sextet antifermion coupling to up-type quarks ($\bar{\Psi}_u$) based on a CMS dijet resonance search at $\sqrt{s}=13\,\text{TeV}$. All limits are computed assuming $\sigma(pp \to \bar{\Psi}_u)$ as displayed in \hyperref[f2]{Figure 3}; but red regions assume unit branching fraction to $\bar{u}g$, while green regions take more realistic branching fractions of around 26\%. In the gray region, where $m_{\Psi_u} > 2\Lambda_{\Psi_u}$, the EFT is unlikely to be consistent.}
\end{figure}

We first examine the single conventional signal with existing constraints and map some experimental results onto our EFT parameter space. In particular, we note that the second line of \eqref{sFmodel} --- which enables single sextet fermion production \emph{via} quark-gluon fusion --- also allows sextet fermions to decay to a quark and a gluon. The full process $pp \to \Psi_q \to \bar{q}g$ (etc.) allows us to constrain the sextet fermions $\Psi_q$ as dijet resonances. Both experimental collaborations have conducted a number of dijet-resonance searches during Run 2 of the LHC. The search easiest to interpret within our model framework was conducted by the CMS collaboration using up to $\mathcal{L} = 36\,\text{fb}^{-1}$ of $pp$ collisions at $\sqrt{s}=13\,\text{TeV}$ \cite{CMS:2018s1}. This analysis targets dijet resonances over a wide mass range ($m_{jj} \in [0.6,8.0]\,\text{TeV}$) and is specifically used to constrain (among others) a benchmark model of excited first-generation quarks decaying to a gluon and a same-flavor quark ($q^* \to qg, q \in \{u,\bar{u},d,\bar{d}\})$. We can use the model-independent limits at 95\% confidence level (CL) \cite{Read:2002cls} on the fiducial cross section $\sigma(pp \to X) \times \text{BF}(X \to qg) \times \mathcal{A}$ computed by CMS to estimate constraints on our sextet fermions decaying to a gluon and a first-generation quark. Our estimates for the fermion that couples to up-type quarks, $\Psi_u$, are displayed in \hyperref[f6]{Figure 6}.

\hyperref[f6]{Figure 6} is in the $(m_{\Psi_u},\Lambda_{\Psi_u})$ plane. We provide no results in the region where $m_{\Psi_u} > 2\Lambda_{\Psi_u}$, since we take $m_{\Psi_u} = 2\Lambda_{\Psi_u}$ as a threshold past which the effective field theory's applicability is dubious. Next, there are two (four) exclusion regions in the plot; solid curves denote observed limits and dashed curves indicate expected limits. We have provided two sets of results in order to show the importance of the branching fraction of the sextet fermion to $\bar{u}g$. The green region(s) correspond closely to the results displayed in Figures \hyperref[f3]{3} and \hyperref[f4]{4}: all couplings and cutoffs except for $\Lambda_{\Psi_u}$ take the same values as in the previous plots. Recall that in these benchmarks, the couplings $\kappa_u^I,\kappa_{uB}^I$ take the same values of $\mathcal{O}(10^{-1})$ for each generation $I$. This choice ultimately produces\footnote{For reference, the other non-negligible branching fractions in this benchmark are roughly as follows: $\text{BF}(\Psi_u \to \bar{c}g) \approx 0.27$, $\text{BF}(\Psi_u \to \bar{t}g) \approx 0.21$, $\text{BF}(\Psi_u \to \bar{u}g A) \approx 0.18$, $\text{BF}(\Psi_u \to \bar{c}gA) \approx 0.06$, and $\text{BF}(\Psi_u \to \bar{t}gA) \approx 0.001$.} $\text{BF}(\Psi_u \to \bar{u}g) \approx 0.26$. We emphasize that these limits apply to the up-type sextet \emph{antifermion} decay, $\bar{\Psi}_u \to ug$, since --- \emph{viz.} \hyperref[f3]{Figure 3} --- the single-production process $pp \to \bar{\Psi}_u$ enjoys the largest cross section at LHC. Similar reinterpretations for the other sextet fermions would yield looser constraints, all else being equal. Finally, the red region(s) use the same cross sections but take $\text{BF}(\Psi_u \to \bar{u}g) = 1$. These choices could be made consistent by appropriately adjusting $\kappa_u^1, \kappa_{uB}^1$ relative to the couplings to heavier up-type quarks. These results serve as a worst-case (high cross section accompanied by high branching fraction) scenario for our sextet fermions from an integrated luminosity of $\mathcal{L} \leq 36\,\text{fb}^{-1}$. In general, we find that light sextet fermions can still be accommodated by these data if $\Lambda_{\Psi_u}$ is in the low multi-TeV range. The limits on this cutoff weaken to nearly $1\,\text{TeV}$ in the ``realistic'' scenario plotted in green.

Dijet-resonance searches are the only searches of which we are aware that currently target a signature produced by these sextet models. Even for these searches, a notable gap exists for sextets decaying to a heavy quark (especially a top quark) and a gluon (again, the CMS search \cite{CMS:2018s1} targets resonances decaying to first-generation quarks). A search tailored to fill this gap may be a good avenue of future study. In the interest of completeness, we note that there exists a CMS search \cite{ext_2018}, using $\mathcal{L} \approx 37\,\text{fb}^{-1}$ of Run 2 data, for \emph{pair-produced} spin-$\tfrac{3}{2}$ color triplets $t^*$ (``excited top quarks'') each decaying promptly to a top quark and a gluon. This search, finding no signal of physics beyond the Standard Model, was used to exclude excited top quarks of around $m_{t^*} = 1\,\text{TeV}$ with fiducial cross sections $\sigma(pp \to t^*\bar{t}^*) \times \text{BF}^2(t^* \to t g)\approx 100\,\text{fb}$. A similar lower limit would be imposed by this search on a pair-produced up-type sextet fermion $m_{\Psi_u}$ given identical acceptances, but a detailed reinterpretation or a dedicated experimental analysis would be required to obtain credible constraints on spin-0 or spin-$\tfrac{1}{2}$ sextets producing this final state. A search for final states involving top quarks from singly produced color sextets would also be welcome.

The other important signals that can be produced by our model catalog are exotic and likely also require dedicated reinterpretations or novel search strategies. The sextet fermions, per the last line of \eqref{sFmodel}, undergo a suppressed decay to a quark, a gluon, and a photon or $Z$ boson. This decay minimally (in the case of single fermion production) produces an interesting dijet resonance + $\gamma/Z$ signature. The sextet scalars, which we have neglected so far, generate another ``dijet resonance-adjacent'' signature by decaying to a quark, a gluon, and a lepton. Finally, we note that the pair production of any sextet --- which occurs copiously at LHC --- potentially generates an array of interesting signatures, particularly for the fermion. For instance, since one fermion could undergo the two-body decay while the other decays to three or even four SM particles, one could expect signatures comprising at least four jets and up to four electrically neutral bosons (and the decays of up to four $Z$ bosons would render these signatures even more complex). Searches for any of these signals would be excellent ways to leverage the higher luminosity of the next planned run of the LHC.

On the other hand, an alternative set of constraints could potentially be imposed on some color-sextet models by searches for flavor-changing neutral currents (FCNCs), which do not exist at tree level in the Standard Model and are very tightly constrained \cite{10.1093/ptep/ptaa104}. It has been observed \cite{PhysRevD.79.015017,PhysRevD.87.115019}, for instance, that color-sextet scalars coupling to quark pairs (which, recall, is a renormalizable interaction; \emph{viz.} \hyperref[t8]{Table VIII}) can generate flavor-changing processes at tree and one-loop level. The strongest limits can be expressed in our notation as
\begin{align}\label{FCNCs}
|\lambda_{11}\lambda^*_{22}| \leq \mathcal{O}(10^{-6})\left(\frac{m_{\Phi}}{\text{TeV}}\right)^2
\end{align}
from tree-level sextet scalar contributions to mixing of neutral kaons ($K^0$-$\bar{K}^0$, with down-type scalars $\Phi_d$) and $D$-mesons ($D^0$-$\bar{D}^0$, with up-type scalars $\Phi_u$)\footnote{Loop contributions yield lighter constraints of $\mathcal{O}(10^{-2})\, m_{\Phi}\, \text{TeV}^{-1}$ or larger.}, and \begin{align}\label{Bdecays}
|\lambda_{23}\lambda^*_{12}| \leq \mathcal{O}(10^{-2})\ \ \ \text{and similar}
\end{align}
from flavor-changing nonleptonic $B$-meson decays mediated at tree level by down-type scalars \cite{PhysRevD.87.115019}.

Similar limits could apply to other models with color sextets. For example, the non-renormalizable sextet fermion model \eqref{sFmodel} we sought to constrain in \hyperref[f6]{Figure 6} can generate meson-mixing diagrams at effective one-loop order. It could therefore be instructive to estimate the size of these FCNC contributions. The superficial degree of ultraviolet (UV) divergence of these box diagrams in $d=4$ dimensions turns out to be
\begin{align}\label{divergence}
D = 4d - N_f - 2N_b + \sum_i h_i N_{v_i} = 2,
\end{align}
where $N_{f,b}$ denote the number of fermion and boson propagators in the loop and the diagram contains $N_{v_i}$ vertices of type $i$ with $h_i$ derivatives each. But the reality is more complicated: these are really multi-loop diagrams, and our ignorance of the UV physics is reflected by a factor of a UV cutoff $\Lambda_{\Psi}^{-1}$ for each of the four vertices forming the boxes. We could therefore estimate the UV divergence of these loops as $D \sim [\text{loop momentum}]^2 \Lambda_{\Psi}^{-4} \to -2$, supposing that the EFT cutoff itself is used to regularize the superficially divergent loop integrals. This degree of divergence matches that of the sextet scalar box contributions to meson mixing discussed above \cite{PhysRevD.87.115019}, so it may indeed be necessary to suppress the quark-sextet couplings $\kappa_q^I$ in this model to evade FCNC constraints\footnote{Such constraints would apply to \emph{all} couplings $\kappa_q^I$, since unlike for tree-level diagrams there is no ``flavor texture'' that will suppress the box diagrams.}. On the other hand, models of sextet scalars or fermions that lack a ``pure'' sextet-quark-quark or sextet-quark-gluon vertex (unaccompanied by leptons or bosons) appear to evade these constraints. But it must be emphasized that this line of thinking is inconclusive without knowledge of the microscopic physics: there may exist diagrams associated with the degrees of freedom that have been integrated out that enhance or interfere with the loops we can compute within the EFT framework. We therefore leave a more detailed investigation of FCNC constraints on our effective operators to future work.
\section{Conclusions}
\label{s4}

In this work we have taken a new look at the possible interactions of beyond-the-Standard Model particles
that are charged under the SM color gauge group $\mathrm{SU}(3)_{\text{c}}$. Such states can be copiously produced at the LHC, and it is important to understand the space of their possible interactions in order to understand how LHC data constrains their existence. In this work, we have explored the gauge-invariant interactions of fields in the six-dimensional (sextet) representation of $\mathrm{SU}(3)_{\text{c}}$, producing a (large) catalog of operators, many of which have received little or no attention in the literature to date but may produce distinct phenomenology worthy of investigation. We have focused on color sextets in this work because they transform in the lowest-dimensional representation of $\mathrm{SU}(3)_{\text{c}}$ not yet observed in Nature. We have specifically focused on higher-dimensional interactions linking color-sextet fermions and scalars to a SM quark and a gluon (and possibly additional particles), computing cross sections for a variety of production modes and surveying existing LHC constraints on these particles. 

Much of what we have done here is intended to set up future work. On one hand, one could undertake a much more thorough investigation of the specific color-sextet models we introduced above; such a study could compute next-to-leading order (NLO) corrections within the EFT framework or could propose an ultraviolet completion for one or more of the operators we consider\footnote{For example, the operator on the second line of \eqref{sFmodel} could straightforwardly be generated by a loop involving SM quarks and a color-triplet scalar, \emph{\`{a} la} squarks.}. It could also be worthwhile to rigorously compute sensitivity projections for any of the non-standard signatures we described in \hyperref[s3.2]{Section III.B} at the high-luminosity LHC. Such projections would be derived from a tailored selection strategy, which would be interesting to develop. It would also be natural to complete the color-sextet operator catalog at mass dimension seven, or to allow for extended sectors comprising sextets with distinct weak hypercharges or non-trivial transformations under $\mathrm{SU}(2)_{\text{L}}$.

On the other hand, as we hopefully have demonstrated with the example of color sextets, the method we have employed to build our operator catalog can be used to build a variety of phenomenological models containing higher representations of $\mathrm{SU}(3)_{\text{c}}$. Similar catalogs could be built for other representations, including the more frequently studied triplets and octets but also higher-dimensional fields; we plan to investigate several such catalogs going forward. The interactions we unearth in these catalogs could also be embedded in richer or more complete theories, for instance \emph{via} the incorporation of a SM gauge singlet as a dark matter candidate.

\appendix

\section{An excursion in SU(3) representation theory}
\label{aA}

The concrete realization of a model incorporating any of the operators cataloged in \hyperref[s2]{Section II} of this work requires explicit knowledge of the gauge-invariant combinations of the relevant fields. While some of our operators contain fields charged under $\mathrm{SU}(2)_{\text{L}}$, all of the exotic gauge singlets we have constructed belong to $\mathrm{SU}(3)_{\text{c}}$. Enumerating the gauge-invariant contractions of a given set of $\mathrm{SU}(3)_{\text{c}}$ multiplets amounts to computing the Clebsch-Gordan coefficients connecting the irreducible (color) representations in which each multiplet transforms. While there exist some works that confront this problem in various contexts \cite{osti_32585,Kaeding_1995,Alex_2011,martins2019su3}, explicit results suitable for fundamental particle physics are difficult to find. In this appendix, we construct the two minimal gauge-invariant combinations of a color sextet with two other color-charged fields. We review and extend some known basis-independent results, and we provide for the first time a new set of Clebsch-Gordan coefficients in a familiar basis well suited for integration into public computing tools.

Let a field --- for definiteness, a Dirac fermion --- in the sextet representation of $\mathrm{SU}(3)_{\text{c}}$ be indexed by $\psi_s$, $s \in \{1,\dots,6\}$. In analogy with a quark $q_i$ transforming in the fundamental representation of $\mathrm{SU}(3)_{\text{c}}$, a lowered index corresponds to the representation in question ($\boldsymbol{6}$), while a raised index (\emph{e.g.} $\bar{\psi}^s$) denotes the conjugate representation ($\boldsymbol{\bar{6}}$). Two of the product decompositions of two irreducible representations of $\mathrm{SU}(3)$ we studied in the body of this work are
\begin{align}\label{A1}
    \boldsymbol{3} \otimes \boldsymbol{3} = \boldsymbol{6} \oplus \boldsymbol{\bar{3}}\ \ \ \text{and}\ \ \ \boldsymbol{3} \otimes \boldsymbol{8} = \boldsymbol{3} \oplus \boldsymbol{\bar{6}} \oplus \boldsymbol{15}.
\end{align}
We noted in \hyperref[s2]{Section II} that these decompositions imply the existence of the \emph{three-field invariants} $\boldsymbol{3}\, \otimes\, \boldsymbol{3}\, \otimes\, \boldsymbol{\bar{6}}$ and $\boldsymbol{3}\, \otimes\, \boldsymbol{6}\, \otimes\, \boldsymbol{8}$. We displayed a number of operators in Tables \hyperref[t8]{VIII} and \hyperref[t8]{IX} based on these invariants that couple sextets to (respectively) quark pairs and a quark and a gluon. While the first family of couplings has received some attention \cite{Han_2010,Han_2010_2}, the latter (to our knowledge) has not. This appendix introduces the explicit group-theoretical objects required for our novel analysis while making contact with known mathematical results.

We work in the basis where the generators of the fundamental ($\boldsymbol{3}$) representation of $\mathrm{SU}(3)$ are proportional to the Gell-Mann matrices: $2\bt{t}_{\boldsymbol{3}}^a = \lambda^a$, $a\in \{1,\dots,8\}$. We take the generators of the adjoint ($\boldsymbol{8}$) representation to be $[\bt{t}_{\boldsymbol{8}}^a]_b^{\ \,c} = -\ii f^{a\  c}_{\ \,b}$, where $f^{abc}$ are the structure constants appearing in the $\mathrm{SU}(3)$ algebra
\begin{align}\label{A2}
    [\bt{t}_{\boldsymbol{3}}^a,\bt{t}_{\boldsymbol{3}}^b] = \ii f^{ab}_{\ \ \,c}\, \bt{t}^c_{\boldsymbol{3}} .
\end{align}
The commutation relations \eqref{A2} are satisfied by the generators of every representation of $\mathrm{SU}(3)$, which are also traceless and Hermitian. A set of eight $6 \times 6$ matrices $\bt{t}_{\boldsymbol{6}}^a$ satisfying these criteria, which are therefore valid generators of the sextet representation in the Gell-Mann basis, is:
\begin{widetext}
\begin{align}\label{gen6}
\nonumber \bt{t}_{\boldsymbol{6}}^1 &= \frac{1}{2}\begin{pmatrix}
0 & \sqrt{2} & 0 & 0 & 0 & 0\\
\sqrt{2} & 0 & \sqrt{2} & 0 & 0 & 0\\
0 & \sqrt{2} & 0 & 0 & 0 & 0\\
0 & 0 & 0 & 0 & 0 & 1\\
0 & 0 & 0 & 0 & 0 & 0\\
0 & 0 & 0 & 1 & 0 & 0\end{pmatrix},\ \ \ \ \ \ \ \ \ \ \ \ \ \ \ \ \ \ \ \bt{t}_{\boldsymbol{6}}^2 = \frac{1}{2}\begin{pmatrix} 0 & 0 & 0 & 0 & 0 & \sqrt{2}\\
0 & 0 & 0 & 1 & 0 & 0\\
0 & 0 & 0 & 0 & 0 & 0\\
0 & 1 & 0 & 0 & 0 & 0\\
0 & 0 & 0 & 0 & 0 & \sqrt{2}\\
\sqrt{2} & 0 & 0 & 0 & \sqrt{2} & 0\end{pmatrix},\\[3ex]
\nonumber \bt{t}_{\boldsymbol{6}}^3 &= \frac{1}{2} \begin{pmatrix}
0 & 0 & 0 & 0 & 0 & 0\\
0 & 0 & 0 & 0 & 0 & 1\\
0 & 0 & 0 & \sqrt{2} & 0 & 0\\
0 & 0 & \sqrt{2} & 0 & \sqrt{2} & 0\\
0 & 0 & 0 & \sqrt{2} & 0 & 0\\
0 & 1 & 0 & 0 & 0 & 0\end{pmatrix},\ \ \ \ \ \ \ \ \ \ \ \ \ \ \ \ \ \ \ \bt{t}_{\boldsymbol{6}}^4 = \frac{1}{2}\begin{pmatrix}
0 & -\ii \sqrt{2} & 0 & 0 & 0 & 0\\
\ii\sqrt{2} & 0 & -\ii \sqrt{2} & 0 & 0 & 0\\
0 & \ii \sqrt{2} & 0 & 0 & 0 & 0\\
0 & 0 & 0 & 0 & 0 & \ii\\
0 & 0 & 0 & 0 & 0 & 0\\
0 & 0 & 0 &-\ii & 0 & 0\end{pmatrix},\\[3ex]
\nonumber \bt{t}_{\boldsymbol{6}}^5 &= \frac{1}{2} \begin{pmatrix}
0 & 0 & 0 & 0 & 0 & -\ii \sqrt{2}\\
0 & 0 & 0 &-\ii & 0 & 0\\
0 & 0 & 0 & 0 & 0 & 0\\
0 & \ii & 0 & 0 & 0 & 0\\
0 & 0 & 0 & 0 & 0 & \ii \sqrt{2}\\
\ii \sqrt{2} & 0 & 0 & 0 & -\ii \sqrt{2} & 0\end{pmatrix},\ \ \ \ \ \ \ \ \ \ \bt{t}_{\boldsymbol{6}}^6 = \frac{1}{2} \begin{pmatrix}
0 & 0 & 0 & 0 & 0 & 0\\
0 & 0 & 0 & 0 & 0 & -\ii\\
0 & 0 & 0 &-\ii \sqrt{2} & 0 & 0\\
0 & 0 &\ii \sqrt{2} & 0 & -\ii \sqrt{2} & 0\\
0 & 0 & 0 & \ii \sqrt{2} & 0 & 0\\
0 & \ii & 0 & 0 & 0 & 0\end{pmatrix},\\[3ex]
\bt{t}_{\boldsymbol{6}}^7 &= \frac{1}{2} \begin{pmatrix}
2 & 0 & 0 & 0 & 0 & 0\\
0 & 0 & 0 & 0 & 0 & 0\\
0 & 0 &-2 & 0 & 0 & 0\\
0 & 0 & 0 &-1 & 0 & 0\\
0 & 0 & 0 & 0 & 0 & 0\\
0 & 0 & 0 & 0 & 0 & 1\end{pmatrix},\ \ \ \ \ \ \ \ \ \ \ \ \ \ \ \ \ \ \ \ \ \ \bt{t}_{\boldsymbol{6}}^8 = \frac{1}{2\sqrt{3}}\begin{pmatrix}
2 & 0 & 0 & 0 & 0 & 0\\
0 & 2 & 0 & 0 & 0 & 0\\
0 & 0 & 2 & 0 & 0 & 0\\
0 & 0 & 0 & -1 & 0 & 0\\
0 & 0 & 0 & 0 & -4 & 0 \\
0 & 0 & 0 & 0 & 0 & -1\end{pmatrix}.
\end{align}
\end{widetext}

It is useful to consider generators of reducible (``product'') representations of $\mathrm{SU}(3)$. The objects relevant to our discussion can be constructed systematically using the generators we have provided above. In particular, the generators $\bt{t}_{\textbf{r}_1 \otimes \textbf{r}_2}^a$ of the direct product of irreducible representations $\textbf{r}_1$ and $\textbf{r}_2$ are given by
\begin{align}\label{A4}
\nonumber    \bt{t}^a_{\textbf{r}_1 \otimes \textbf{r}_2} &= \bt{t}^a_{\textbf{r}_1} \oplus \bt{t}^a_{\textbf{r}_2}\\
    &\equiv \bt{t}^a_{\textbf{r}_1} \otimes \bt{1}_{\textbf{r}_2} + \bt{1}_{\textbf{r}_1} \otimes \bt{t}^a_{\textbf{r}_2},
\end{align}
where the \emph{Kronecker sum} $\oplus$ is defined in terms of the \emph{Kronecker products} (suggestively denoted by the same symbol $\otimes$ as the direct product of representations) between the generators of $\{\textbf{r}_1, \textbf{r}_2\}$ and the identity matrices with the dimensions of $\{\textbf{r}_2,\textbf{r}_1\}$. In order to elucidate this point, and to make contact with more familiar notation, we rewrite the last line of \eqref{A4} as
\begin{align}
    [\bt{t}^a_{\textbf{r}_1 \otimes \textbf{r}_2}]_{I_1 I_2}^{\ \ \ \ \, J_1 J_2} = [\bt{t}^a_{\textbf{r}_1}]_{I_1}^{\ \ J_1} \delta_{I_2}^{\ \ J_2} + \delta_{I_1}^{\ \ J_1} [\bt{t}^a_{\textbf{r}_2}]_{I_2}^{\ \ J_2}
\end{align}
with $\{I_i,J_i\}$ indexing the representation $\textbf{r}_i$, $i \in \{1,2\}$. The resulting generators $\bt{t}^a_{\textbf{r}_1\otimes \textbf{r}_2}$ are again traceless and Hermitian, but now of dimension $\dim \textbf{r}_1 \times \dim \textbf{r}_2$. The operation \eqref{A4} can be iterated upon, so for instance the generators of a direct product of three representations are given by
\begin{align}
\nonumber \bt{t}^a_{\textbf{r}_1 \otimes \textbf{r}_2 \otimes \textbf{r}_3} = \bigoplus_{i=1}^3 \bt{t}^a_{\textbf{r}_i} &= \bt{t}^a_{\textbf{r}_1} \oplus \bt{t}^a_{\textbf{r}_2} \oplus \bt{t}^a_{\textbf{r}_3} \\ &= (\bt{t}^a_{\textbf{r}_1 \otimes \textbf{r}_2}) \oplus \bt{t}^a_{\textbf{r}_3}.
\end{align}
We provide all of this exposition because any gauge-invariant linear combination $\mathcal{I}_{\textbf{r}_1 \otimes \cdots \otimes \textbf{r}_N}$ of $M$ fields $\psi^j_{I_i}$ in $N \leq M$ (not necessarily distinct) irreducible representations $\{\textbf{r}_1, \dots, \textbf{r}_N\}$ of $\mathrm{SU}(3)$ (or any semisimple Lie group), which can be written modulo index height as \cite{fonseca2013renormalization}
\begin{align}\label{A7}
    \mathcal{I}_{\textbf{r}_1 \otimes \cdots \otimes \textbf{r}_N} = \kappa^{I_1, \dots,I_N} \psi^1_{I_1} \dots \psi^M_{I_N}
\end{align}
with summation understood over repeated indices, must satisfy
\begin{align}\label{A8}
\kappa^{I_1,\dots,I_N} [\bt{t}^a_{\textbf{r}_1 \otimes \cdots \otimes \textbf{r}_N}]_{I_1, \dots, I_N}^{\ \ \ \ \ \ \ \ \ J_1, \dots, J_N} = 0\ \ \ \forall\, \{J_i,a\}.
\end{align}
The coefficients $\kappa^{I_1,\dots, I_N}$ appearing in the linear combinations \eqref{A7}, \eqref{A8} are closely related to the Clebsch-Gordan coefficients connecting the $N$ irreducible representations. The coefficients we want simply have to be extracted and interpreted in a specific way. \eqref{A8} implies that these coefficients can be systematically computed in any basis by finding the kernel of the (potentially large) matrix
\begin{align}\label{A9}
\bt{M}_{\textbf{r}_1 \otimes \cdots \otimes \textbf{r}_N} = \begin{pmatrix}
[\bt{t}^1_{\textbf{r}_1 \otimes \cdots \otimes \textbf{r}_N}]^{\transpose}\\
\vdots\\
[\bt{t}^a_{\textbf{r}_1 \otimes \cdots \otimes \textbf{r}_N}]^{\transpose}\end{pmatrix}.
\end{align}
The number of independent gauge-invariant combinations of the $N$ fields is given by $\dim \ker \bt{M}_{\textbf{r}_1 \otimes \cdots \otimes \textbf{r}_N}$. The elements of the $(\dim \textbf{r}_1 \times \cdots \times \dim \textbf{r}_N) \times 1$ vectors in $\ker \bt{M}_{\textbf{r}_1 \otimes \cdots \otimes \textbf{r}_N}$ are (up to a global factor) the desired Clebsch-Gordan coefficients. Per our notation in \eqref{A8}, these are read off in an order determined by the construction of the product-representation generators. As an intuitive example, the first two non-vanishing elements (in the Gell-Mann basis) of the only vector in $\ker \bt{M}_{\boldsymbol{6} \otimes \boldsymbol{3} \otimes \boldsymbol{8}}$ in $\mathrm{SU}(3)$ --- which is clearly relevant to our phenomenological study of color sextets --- are elements $(10,1)$ and $(13,1)$. These we label as $\bt{J}^{\,1,2,2}$ and $\bt{J}^{\,1,2,5}$, elements of the first of six $3 \times 8$ matrices $\bt{J}^{\,s\, ia}$ ($s\in \{1,\dots,6\}$, $i \in \{1,2,3\}$, $a \in \{1,\dots,8\}$). This is essentially the method employed by the recently published \textsc{Mathematica}$^\copyright$\ \cite{Mathematica} package \textsc{GroupMath}, which however works in what is sometimes called the \emph{Chevalley-Serre basis} \cite{Fonseca_2021}. We have performed these calculations in the Gell-Mann basis in order to obtain basis-dependent results compatible with the literature and the popular computer tools \textsc{FeynRules} and \textsc{MadGraph5\texttt{\textunderscore}aMC@NLO} \cite{FR_OG,FR_2,MG5,MG5_EW_NLO}.

It should be noted that the method described above in fact yields \emph{some multiple} of the Clebsch-Gordan coefficients $\bt{C}^{I_1,\dots,I_N}$ for any invariant combination of fields in compatible irreducible representations of a given group. If the coefficients are subsequently normalized to satisfy 
\begin{align}\label{A10}
\tr \bt{C}^{I_1}\, \bar{\!\bt{C}}_{J_1} = \delta^{I_1}_{\ \ J_1}\ \ \ \text{with}\ \ \ \bar{\!\bt{C}}_{I_1,\dots,I_N} \equiv [\bt{C}^{I_1, I_N, I_{N-1},\dots,I_2}]^*,
\end{align}
where the trace is performed over all subleading indices $\{I_2,\dots,I_N; J_2,\dots,J_N\}$, then the coefficients further satisfy a relation of the form
\begin{multline}\label{A11}
[\bt{t}_{\boldsymbol{\bar{\textbf{r}}}_1}^a]^{I_1}_{\ \ J_1} = \bt{C}^{I_1,\dots, I_N} \, \bar{\!\bt{C}}_{J_1,\dots, J_N}\\ \times [\bt{t}^a_{\textbf{r}_2 \otimes \cdots \otimes \textbf{r}_N}]_{I_2,\dots,I_N}^{\ \ \ \ \ \ \ \ \ J_2,\dots, J_N},
\end{multline}
which allows one to extract generators of a given representation in an invariant combination from the generators of the direct product of all the \emph{other} representations present in that invariant. Note that the generators on the left-hand side of \eqref{A11} are those of the conjugate representation $\boldsymbol{\bar{\textbf{r}}}_1$; recall that the generators of $\textbf{r}_1$ can then be recovered according to $[\bt{t}^a_{\textbf{r}_1}]_{I_1}^{\ \ J_1} = -\{[\bt{t}^a_{\boldsymbol{\bar{\textbf{r}}}_1}]^{I_1}_{\ \ J_1}\}^*$.

We finally provide explicit results relevant to the body of this work; \emph{i.e.}, to the gauge-invariant combinations (three-field invariants) $\boldsymbol{3}\, \otimes\, \boldsymbol{3}\, \otimes\, \boldsymbol{\bar{6}}$ and $\boldsymbol{3}\, \otimes\, \boldsymbol{6}\, \otimes\, \boldsymbol{8}$ of $\mathrm{SU}(3)_{\text{c}}$ implied by the decompositions \eqref{A1}. In particular, we recoup the generators $\bt{t}_{\boldsymbol{6}}$ of the sextet representation of $\mathrm{SU}(3)$, given in the Gell-Mann basis by \eqref{gen6}, according to
\begin{align}
    [\bt{t}_{\boldsymbol{6}}^a]_s^{\ \,t} &= \bt{K}_s^{\ \,ij} \bar{\bt{K}}{}^t_{\ \,lk}\, [\bt{t}^a_{\boldsymbol{3}\otimes \boldsymbol{3}}]_{ij}^{\ \ \,kl}\\
    &= -\{\bt{J}^{\,s\, ib}\, \bar{\bt{J}}{}_{t\,cj}\, [\bt{t}^a_{\boldsymbol{3}\otimes \boldsymbol{8}}]_{ib}^{\ \ \,jc}\}^*,
\end{align}
with
\begin{widetext}
\begin{align}\label{Kdef}
\nonumber \bt{K}_{1} &= \begin{pmatrix}
1 & 0 & 0\\
0 & 0 & 0\\
0 & 0 & 0\end{pmatrix},\ \ \ \bt{K}_{2} = 
\frac{1}{\sqrt{2}}\begin{pmatrix} 0 & 1 & 0\\
1 & 0 & 0\\
0 & 0 & 0\end{pmatrix},\ \ \ \bt{K}_3 = \begin{pmatrix} 0 & 0 & 0\\
0 & 1 & 0\\
0 & 0 & 0\end{pmatrix},\\[3ex] \bt{K}_4 &= \frac{1}{\sqrt{2}} \begin{pmatrix} 0 & 0 & 0\\
0 & 0 & 1\\
0 & 1 & 0\end{pmatrix},\ \ \ \bt{K}_5 = \begin{pmatrix} 0 & 0 & 0\\
0 & 0 & 0\\
0 & 0 & 1\end{pmatrix},\ \ \ \bt{K}_6 = \frac{1}{\sqrt{2}} \begin{pmatrix}0 & 0 & 1\\
0 & 0 & 0\\
1 & 0 & 0\end{pmatrix}
\end{align}
\end{widetext}
and
\begin{widetext}
\begin{align}\label{Jdef}
\nonumber \bt{J}^{\,1} &= \frac{1}{2}\begin{pmatrix}
0 & 0 & 0 & 0 & 0 & 0 & 0 & 0\\
0 & -\ii & 0 & 0 & 1 & 0 & 0 & 0\\
\ii & 0 & 0 & -1 & 0 & 0 & 0 & 0
\end{pmatrix},\ \ \ \ \ \ \ \ \ \ \ \ \ \bt{J}^{\,2} = \frac{1}{4} \begin{pmatrix}
0 & \ii\sqrt{2} & 0 & 0 & -\sqrt{2} & 0 & 0 & 0\\
0 & 0 & -\ii\sqrt{2} & 0 & 0 & \sqrt{2} & 0 & 0\\
0 & 0 & 0 & 0 & 0 & 0 & -2\ii\sqrt{2} & 0\end{pmatrix},\\[3ex]
\nonumber \bt{J}^{\,3} &= \frac{1}{2}\begin{pmatrix}
0 & 0 & \ii & 0 & 0 & -1 & 0 & 0\\
0 & 0 & 0 & 0 & 0 & 0 & 0 & 0\\
-\ii & 0 & 0 & -1 & 0 & 0 & 0 & 0\end{pmatrix},\ \ \ \ \ \ \ \ \ \ \, \bt{J}^{\,4} = \frac{1}{4}\begin{pmatrix}
0 & 0 & 0 & 0 & 0 & 0 & \ii\sqrt{2} & -\ii\sqrt{6}\\
\ii\sqrt{2} & 0 & 0 & \sqrt{2} & 0 & 0 & 0 & 0\\
0 & -\ii\sqrt{2} & 0 & 0 & -\sqrt{2} & 0 & 0 & 0 \end{pmatrix},\\[3ex]
\bt{J}^{\,5} &= \frac{1}{2}\begin{pmatrix}
0 & 0 & -\ii & 0 & 0 & -1 & 0 & 0\\
0 & \ii & 0 & 0 & 1 & 0 & 0 & 0\\
0 & 0 & 0 & 0 & 0 & 0 & 0 & 0 \end{pmatrix},\ \ \ \ \ \ \ \ \ \ \ \ \ \bt{J}^{\,6} = \frac{1}{4} \begin{pmatrix}
-\ii\sqrt{2} & 0 & 0 & \sqrt{2} & 0 & 0 & 0 & 0\\
0 & 0 & 0 & 0 & 0 & 0 & \ii\sqrt{2} & \ii\sqrt{6}\\
0 & 0 & \ii\sqrt{2} & 0 & 0 & \sqrt{2} & 0 & 0\end{pmatrix}.
\end{align}
\end{widetext}
The first set \eqref{Kdef} of Clebsch-Gordan coefficients, which connect the $\boldsymbol{6}$ to the $\boldsymbol{3}\,\otimes\,\boldsymbol{3}$ of $\mathrm{SU}(3)$, are exactly what were computed in the only group-theoretical discussion similar to this appendix of which we are aware \cite{Han_2010}. The second set, \eqref{Jdef}, which connect the $\boldsymbol{\bar{6}}$ to the $\boldsymbol{3}\,\otimes\,\boldsymbol{8}$, have not been published before as far as we know. Both sets of coefficients are similarly normalized ---
\begin{align}\label{A16}
\tr \bt{K}_s \bar{\bt{K}}{}^t = \delta_s^{\ \,t}\ \ \ \text{and}\ \ \ \tr \bt{J}_s\, \bar{\!\bt{J}}{}^{\,t} = \delta_s^{\ \,t}
\end{align}
--- and satisfy 
\begin{align}
    \bar{\bt{K}}{}^s_{\ \,ji} = [\bt{K}_s^{\ \,ji}]^{\dagger}= \bt{K}_s^{\ \,ij}\ \ \ \text{and}\ \ \ \bar{\bt{J}}{}_{s\, ai} = [\bt{J}^{\,s\, ia}]^{\dagger}.
\end{align}

Our final remark concerns the relationship between the two sets of Clebsch-Gordan coefficients $\bt{K}_s^{\ \,ij}$ and $\bt{J}^{\,s\, ia}$. We find that the latter set can be constructed using a particular combination of the former set with other group-theoretical objects. In particular, we have that
\begin{multline}\label{A18}
    \bt{J}^{\,s\, ia} = -\ii \sqrt{2}\, \bt{L}^{ijk} [\bt{t}_{\boldsymbol{3}}^a]_j^{\ \,l} \bar{\bt{K}}{}^s_{\ \,lk}\\ \text{and}\ \ \ \bar{\bt{J}}{}_{s\, ai} = \ii \sqrt{2}\, \bt{K}_s^{\ \,kl}[\bt{t}_{\boldsymbol{3}}^a]_l^{\ \, j} \bar{\bt{L}}_{\,ijk},
\end{multline}
where $\bt{L}^{ijk}$ are the Clebsch-Gordan coefficients governing the gauge-invariant contraction of three $\mathrm{SU}(3)$ triplets, which is well known to be totally antisymmetric\footnote{In the Gell-Mann basis, these coefficients are proportional to the Levi-Civita symbol: $\sqrt{2}\,\bt{L}^{ijk} = \epsilon^{ijk}$ and $\sqrt{2}\,\bar{\bt{L}}_{\,ijk} = \epsilon_{ijk}$.}. We mention the relations \eqref{A18} because the popular model-building and Monte Carlo simulation tools \textsc{FeynRules} and \textsc{MadGraph5\texttt{\textunderscore}aMC@NLO} have for some time now handled color-sextet fields interacting with quark pairs by defining the Clebsch-Gordan coefficients $\bt{K}_s^{\ \,ij}$ and $\bt{L}^{ijk}$ in terms of (anti-)symmetric combinations of two QCD triplets. Therefore, whereas the ability to directly handle the new coefficients $\bt{J}^{\,s\,ia}$ --- given by \eqref{Jdef} --- would require some significant additions to both public codes, we are able to construct our novel interactions with suitable combinations of existing semi-hard-coded objects. This strategy does not necessarily work for color-sextet interactions with higher-dimensional QCD multiplets, and may in fact be a unique exploit.

\acknowledgments
T.T. is grateful for conversations with Rohini Godbole, Kirtimaan Mohan, and Daniel Whiteson. L.M.C. and T.M. are supported in part by the United States Department of Energy under grant DE-SC0011726. T.T. is supported in part by the United States National Science Foundation under grant no. PHY-1915005.

\bibliographystyle{apsrev4-2}
\bibliography{Bibliography/bibliography}

\end{document}